\documentclass[aps,prb,twocolumn,superscriptaddress,amsmath,amssymb]{revtex4}

\usepackage{graphicx}
\usepackage{dcolumn}
\usepackage{bm}

\begin{document}


\title{Non-Equilibrium Kondo Model with Voltage Bias in a Magnetic Field}

\author{P. Fritsch}
\email{peter.fritsch@physik.lmu.de}
\author{S. Kehrein}
\affiliation{
Physics Department, Arnold Sommerfeld Center for Theoretical Physics, and Center for NanoScience, \\
Ludwig-Maximilians-Universit\"at, Theresienstrasse 37, 80333 Munich, Germany
}

\date{\today}

\begin{abstract}
We derive a consistent 2-loop scaling picture for a Kondo dot in both equilibrium and non-equilibrium situations using the flow equation method.
Our analysis incorporates the important decoherence effects from both thermal and non-equilibrium noise
in a common setting.
We calculate the spin-spin correlation function, the T-matrix, and the magnetization as functions of applied magnetic field,
dc-voltage bias and temperature. In all these quantities we observe characteristic non-equilibrium features for a
nonvanishing external voltage bias like Kondo splitting and strongly enhanced logarithmic corrections.
\end{abstract}

\maketitle

\section{Introduction}

\subsection{Motivation}
The Kondo effect was first observed in the 1930's while measuring the resistivity of ``pure'' metals.
Upon lowering the temperature one finds a minimum in the resistivity of nonmagnetic metals containing a small concentration of magnetic impurities.
When lowering the temperature even further the resistivity increases and saturates at a finite value at zero temperature.
Systematic experimental and theoretical analysis showed that this effect is due to a screening of the impurity spin
by resonant scattering of conduction band electrons leading to an enhanced electron density around the impurities.
Bypassing electrons scatter off these so called spin compensation clouds leading to an enhancement of the resistivity.
The Kondo model has become a paradigm model for strong-coupling impurity physics in condensed matter theory\cite{Hewson,RevModPhys.47.773}.
It has been solved exactly using the Bethe Ansatz\cite{RevModPhys.55.331,tsvelick_1983},
however, dynamical quantities like the impurity spectral function
are not easily accessible within this framework. Many other numerical and analytical methods have been developed since that
can get around this
limitation\cite{RevModPhys.47.773,PhysRevLett.85.1504,bulla:395,PhysRevB.68.014430,LoganDickens,Dickens2001,PhysRevLett.76.1683,PhysRevA.41.5383,PhysRevB.63.140402,PhysRevLett.85.1722}.

Experiments on quantum dots in the Coulomb blockade regime have revived the interest in Kondo physics
\cite{gg_1998,SaraM.Cronenwett07241998,klitzing_1998}.
If the quantum dot is tuned in such a way
that it carries a net spin, resonant tunneling leads to an increase of the conductance up to the unitary limit\cite{GlazmanRaikh,PhysRevLett.61.1768,Wiel09222000}.
For small dc-voltage bias $V \ll T_K$ the system can be described using linear response theory. However, for
$V \gtrsim T_{K}$ linear response
theory starting from the equilibrium ground state is no longer applicable.

In this paper we study a quantum dot in the Kondo regime (Kondo dot) with an applied magnetic field $h$ in the regime
$\text{max}(V,|h|,T)\gg T_K$, where $V$ is the dc-voltage bias and $T$ the temperature.
We diagonalize the Hamiltonian using infinitesimal unitary transformations (flow equations) \cite{wegner1994,kehreinbook}.
Unlike in conventional scaling approaches the high energy states are not integrated out,
instead the states are successively decoupled  from large to small energy differences.
Since all energy conserving processes are retained, the steady current across the dot turns out to be included in the scaling picture.
This current generates a decoherence rate $\Gamma$ that cuts off the logarithmic divergences arising in the Kondo problem,
thereby making the situation $\text{max}(V,|h|,T)\gg T_K$ a weak-coupling problem.
Previous renormalization group (RG) calculations\cite{PhysRevLett.83.384,PhysRevB.62.8154,PhysRevLett.86.4088,Rosch2005,PhysRevLett.90.076804,PhysRevB.70.155301,paaske:155330,PhysRevLett.87.156802}
already established that decoherence effects due to spin relaxation processes play a key role in non-equilibrium.
This was confirmed by a flow equation analysis of the Kondo model with voltage bias\cite{kehrein:056602,kehreinbook,kehrein2007}.
Other new scaling approaches to non-equilibrium problems like the real time renormalization 
group\cite{schoellerRG,jakobs:150603,Korb2007,Schoeller2009}
and the Coulomb gas representation\cite{mitra:085342,segal:195316} are consistent with this general picture and have
added further insights.
At this point one should also mention other new approaches
like the scattering state numerical renormalization group\cite{anders:066804}, the time-dependent density 
renormalization group\cite{boulat:140601}, the scattering state Bethe Ansatz\cite{mehta:216802,mehta2}
and $1/N$-expansion techniques\cite{Aditi2009}
that open up the possibility to describe the very challenging crossover regime for intermediate voltage bias
$V\sim T_{K}$.

In this paper we generalize the flow equation analysis\cite{kehrein:056602,kehreinbook,kehrein2007} to include a magnetic field.
A similar two-loop calculation based on the real time renormalization
group was recently also performed in Ref.~\cite{Reininghaus2009}.
As main results we derive the spin-spin correlation function, the T-matrix and the magnetization
in both equilibrium and non-equilibrium situations. Our results for the non-equilibrium static spin susceptibility~$\chi_{0}(T,V)$ 
at zero external magnetic field $h=0$ were already published in Ref.\cite{kehrein2007}

Let us first have a closer look at the magnetization. 
The equilibrium magnetization is well known from the Bethe Ansatz\cite{RevModPhys.55.331,tsvelick_1983}.
Previous non-equilibrium perturbation theory calculations\cite{PhysRevB.66.085315,paaske:155330,Reininghaus2009}
for the magnetization derived the correct high voltage/temperature $V,T\gg |h|$ behavior, but important logarithmic corrections
in non-equilibrium are missing.
Using the flow equation approach up to two-loop order we will be able to calculate the magnetization including its leading logarithmic corrections
consistently in the whole weak-coupling regime. 

The T-matrix and the closely related impurity spectral function are also well studied
objects\cite{PhysRevLett.85.1504,LoganDickens,Dickens2001,PhysRevB.68.014430,PhysRevLett.85.1722,PhysRevB.70.155301}.
Nevertheless, some parameter regimes like combinations of magnetic field with nonzero voltage bias have not yet been investigated.
We rederive the previous results and give additional insights into the crossover regimes.

The equilibrium spin-spin correlation function is known in all parameter regimes,\cite{PhysRevA.41.5383,PhysRevLett.76.1683,PhysRevB.70.155301,PhysRevLett.91.207203} 
especially in the setting of the equilibrium spin boson model.
We generalize our previous results\cite{kehreinbook,kehrein2007} in both equilibrium and non-equilibrium to include nonzero magnetic fields.
In addition we discuss in detail the interplay of the different decoherence sources on the spin dynamics.\\

The paper is organized as follows.
In Sects.~\ref{Hamiltonian_sec} and \ref{Flow_Equations_sec} we define the model and give a short introduction to the flow equation method.
The flow equations for the Hamiltonian and their scaling analysis are derived in Sect.~\ref{Flow_Ham_sec} (with additional
details in appendices \ref{Norm_Order} and \ref{Trafo_H}). The transformation of the spin operator and the resulting correlation functions
are shown in Sect.~\ref{Dot_spin_sec} and Appendix \ref{XYZ}. In Sect.~\ref{Magnet_sec} we analytically
derive the equilibrium zero temperature magnetization in leading order directly from the transformation of the spin operator.
The calculation of the T-matrix is shown in Sect.~\ref{T-Matrix_sec}.
Numerical results that show similarities and differences between voltage bias and temperature are discussed in Sect.~\ref{V_vs_T_sec}.

\subsection{Non-Equilibrium Kondo Model}\label{Hamiltonian_sec}
The Hamiltonian of a spin-$1/2$ Kondo dot in a magnetic field coupled to two leads is given by
\begin{eqnarray}
H & = & \sum\limits_{p,\alpha,\sigma} (\epsilon_p-\mu_\alpha) c_{p\alpha\sigma}^\dagger c_{p\alpha\sigma} -hS^z \label{H_ini}\\
& \ & + \sum\limits_{p,q,\alpha,\beta} \frac{J_{\alpha\beta}}{2} \left( 
\left( c_{p\alpha\uparrow}^{\dagger}c_{q\beta\uparrow} -  c_{p\alpha\downarrow}^{\dagger}c_{q\beta\downarrow} \right) S^z
\right.\nonumber\\
& \ & \; \left. + \left(  c_{p\alpha\uparrow}^{\dagger}c_{q\beta\downarrow} S^- + \mbox{h.c.} \right) \right) \nonumber
\end{eqnarray}
where $\vec S$ is the impurity spin, $\alpha,\beta=l,r$ label the leads, $\sigma=\uparrow,\downarrow$ is the spin index,
$\mu_{l,r}=\pm V/2$ is the chemical potential,
and $h$ is the magnetic field. Without loss of generality we assume $V\ge0$. 
We are always interested in the isotropic Kondo model as is relevant in quantum dot physics, though most of our calculations can easily
be generalized to the anisotropic case. \\

Analogous to our previous calculations\cite{kehrein:056602,kehreinbook,kehrein2007}
we split the operator space in even and odd combinations of fermionic operators
from the left and right lead:
\begin{eqnarray}
f_{p\sigma} & = & \frac{1}{\sqrt{1+R}} c_{pr\sigma} + \frac{1}{\sqrt{1+R^{-1}}} c_{pl\sigma}
\label{definef} \\
g_{p\sigma} & = & \frac{1}{\sqrt{1+R^{-1}}} c_{pr\sigma} - \frac{1}{\sqrt{1+R}} c_{pl\sigma}\ , \nonumber
\end{eqnarray}
where $R=J_{ll}/J_{rr}$. Note that the $f$- and $g$-operators obey fermionic anticommutation relations.
If the Hamiltonian (\ref{H_ini}) is derived from an underlying Anderson impurity model
\cite{PhysRevLett.83.384,PhysRevB.62.8154}, 
the antisymmetric operators $g_{p\sigma}^\dagger ,g_{p\sigma}$ decouple completely from the dot and the Hamiltonian (\ref{H_ini}) can be
written in terms of the $f-$operators only:
\begin{eqnarray}
H & = & \sum\limits_{p,\sigma} \epsilon_p f_{p\sigma}^\dagger f_{p\sigma} -hS^z \label{H_f_no_normal}\\
& \ & + \sum\limits_{p,q} \frac{J}{2} \left( 
\left( f_{p\uparrow}^{\dagger}f_{q\uparrow} -  f_{p\downarrow}^{\dagger}f_{q\downarrow} \right) S^z
\right.\nonumber\\
& \ & \; \left. + \left(  f_{p\uparrow}^{\dagger}f_{q\downarrow} S^- + \mbox{h.c.} \right) \right)\ , \nonumber
\end{eqnarray}
where $J\stackrel{\rm def}{=}J_{ll}+J_{rr}$ and we have used $J_{lr}^2=J_{rl}^2=J_{ll}J_{rr}$.\cite{PhysRevLett.83.384,PhysRevB.62.8154} 
The Hamiltonian (\ref{H_f_no_normal}) looks formally like a standard Kondo impurity coupled to a conduction band, 
the only difference being the non-equilibrium occupation number distribution of the initial state derived from
(\ref{definef}):
\begin{equation}
\label{defnfp}
n_f(p)=\langle f_{p\sigma}^\dagger f_{p\sigma} \rangle = \left\{
\begin{array}{cl}
0 & ,\ \epsilon_p > V/2 \\
\frac{1}{1+1/R} & ,\ |\epsilon_p| \le V/2 \\
1 & ,\ \epsilon_p< - V/2
\end{array}
\right.
\end{equation}
In equilibrium the Kondo temperature is given by $T_K=D\sqrt{\rho J}\exp(-1/(\rho J))$, where $2D$ is the bandwidth
and $\rho$ the conduction electron density of states (we assume a constant density of states). 
We will use this definition of the Kondo temperature in the remainder of this paper.
For convenience we also set $\rho=1$ in the following.
By using the Hamiltonian (\ref{H_f_no_normal})
we will be able to describe the equilibrium and the non-equilibrium system in a unified scaling picture.
For later reference let us already quote the result for the steady state current in the large dc-voltage 
limit for vanishing external magnetic field\cite{PhysRevLett.83.384,PhysRevB.62.8154}
\begin{equation}
I=\frac{3\pi}{4}\,\frac{1}{(1+R)(1+R^{-1})}\,\frac{V}{\ln^{2}(V/T_{K})} \ .
\label{currentI}
\end{equation}

\subsection{Flow Equations}\label{Flow_Equations_sec}
The flow equation method\cite{kehreinbook,wegner1994} provides a framework to diagonalize a Hamiltonian using
infinitesimal unitary transformations. These are constructed using the differential equation
\begin{equation}
\frac{dH(B)}{dB} = [ \eta(B), H(B) ]\ ,
\end{equation}
where the generator $\eta(B)$ is a suitable antihermitian operator.
$H(B=0)$ is the initial Hamiltonian and
$H(B=\infty)$ the diagonal one.
The generic choice for the generator is given
by the commutator $\eta(B)=[H_0(B),H_{\text{int}}(B)]$, where $H_0(B)$ is the diagonal part of the Hamiltonian and $H_{\text{int}}(B)$
the interaction part.
With this definition of the generator one can define an energy scale $\Lambda_{\text{feq}}=B^{-1/2}$ that corresponds to
the remaining effective bandwidth: Interaction matrix elements with high energy transfer $|\Delta\epsilon|\gtrsim\Lambda_{\text{feq}}$
are eliminated in the Hamiltonian $H(\Lambda_{\text{feq}})$, while processes with smaller energy transfer are still retained.\\

For generic many particle problems the flow generates new interactions,
which appear in higher order of the interaction parameter. To keep track of
the latter we introduce a parameter $\lambda=1$ in the Hamiltonian $H=H_0+\lambda H_{\text{int}}$.
We will only take terms into account that couple back into the flow of the original Hamiltonian up to a certain power
of $\lambda$. This corresponds to a loop expansion in renormalization theory: a n-loop calculation takes terms of order
$\lambda^{n+1}$ into account. We use normal ordering to expand operator products, see
Appendix \ref{Norm_Order} for more details.\\

To evaluate expectation values the operators have to be transformed into the diagonal ($B=\infty$) basis.
Any linear operator $O$ is transformed using
\begin{equation}
\frac{dO(B)}{dB} = [ \eta(B), O(B) ]\ .
\end{equation}
A generic operator will typically generate an infinite number of higher order terms and one has to choose a suitable approximation
scheme, which is again perturbative in the running coupling.

The flow equation approach has been successfully applied to various equilibrium many-body problems,
like dissipative quantum systems\cite{kehrein_mielke,PhysRevB.70.014516},
the two-dimensional Hubbard model\cite{grote,PhysRevB.66.094516},
low-dimensional spin systems\cite{PhysRevLett.87.167204,knetter},
and strong coupling models like the sine-Gordon model\cite{PhysRevLett.83.4914,Kehrein2001512}
and the Kondo model\cite{PhysRevB.63.140402,PhysRevB.69.214413}.
It has also been successfully applied to numerous non-equilibrium initial state problems\cite{hackl:092303,0953-8984-21-1-015601,moeckel:175702,PhysRevB.71.193303}.

\section{Flow of the Hamiltonian}\label{Flow_Ham_sec}

\subsection{Ansatz and Generator}\label{2_loop}
\begin{widetext}
In the following we derive the Hamiltonian flow for the Kondo Hamiltonian (\ref{H_f_no_normal}).
We use the ansatz
\begin{eqnarray}
H_0 & = & \sum\limits_{p,\sigma} \epsilon_p :f_{p\sigma}^\dagger f_{p\sigma}: -h(B)S^z \label{Ansatz}\\
H_{\text{int}} & = & \frac{1}{2} \sum\limits_{p,q} \left( J^{\uparrow}_{pq}(B) :f_{p\uparrow}^\dagger f_{q\uparrow}: 
 - J^{\downarrow}_{pq}(B) :f_{p\downarrow}^\dagger f_{q\downarrow}: \right) S^z 
  + \frac{1}{2} \sum\limits_{p,q} J^\perp_{pq}(B) ( :f_{p\uparrow}^\dagger f_{q\downarrow}: S^- +
:f_{q\downarrow}^\dagger f_{p\uparrow}: S^+ ) \nonumber\\
& \ & +\sum\limits_{p,q,r,s} K^\uparrow_{pq,rs} (B)
( :f_{p\uparrow}^\dagger f_{q\downarrow} f_{r\uparrow}^\dagger f_{s\uparrow} : S^- +
\text{h.c.}) 
+ \sum\limits_{p,q,r,s} K^\downarrow_{pq,rs}(B)
( :f_{p\uparrow}^\dagger f_{q\downarrow} f_{r\downarrow}^\dagger f_{s\downarrow} : S^- + \text{h.c.} ) \nonumber\\
& \ & + \sum\limits_{p,q,r,s} K^\perp_{pq,rs}(B)
:f_{p\uparrow}^\dagger f_{q\downarrow} f_{r\downarrow}^\dagger f_{s\uparrow} : S^z\ , \nonumber
\end{eqnarray}
where $:\;:$ denotes normal ordering with respect to the system without Kondo impurity\cite{note_normal_order},
$K^{\uparrow/\downarrow/\perp}_{pq,rs}(B=0)=0$ and
$J^{\uparrow/\downarrow/\perp}_{pq}(B=0)=J$.
For zero magnetic field the relations
$h=0$, $J^{\uparrow}_{pq}=J^{\downarrow}_{pq}=J^{\perp}_{pq}=J^{\perp}_{qp}$, and $K^\uparrow_{pq,rs}=-K^\downarrow_{pq,rs}=-K^\perp_{pq,rs}/2$
are fulfilled during the flow. The relations $J^{\uparrow}_{pq}=J^{\uparrow}_{qp}$, $J^{\downarrow}_{pq}=J^{\downarrow}_{qp}$,
and $K^\perp_{pq,rs}=K^\perp_{sr,qp}$
are always fulfilled due to hermiticity.
An additionally generated
potential scattering term is neglected since it has no influence on the universal low energy properties of the model\cite{kehrein:056602,kehreinbook,kehrein2007}.
We also drop an uninteresting constant in the flow of the Hamiltonian.
The straightforward derivation of the commutation relations yields the generator
\begin{eqnarray}
\eta(B) & = & \frac{1}{2} \sum\limits_{p,q} (\epsilon_p-\epsilon_q) \left( J^{\uparrow}_{pq}(B) :f_{p\uparrow}^\dagger f_{q\uparrow}:-
J^{\downarrow}_{pq}(B) :f_{p\downarrow}^\dagger f_{q\downarrow}:\right)S^z  \label{Generator}\\
& \ & +  \frac{1}{2} \sum\limits_{p,q}  (\epsilon_p-\epsilon_q +h(B)) J^\perp_{pq}(B) \left( :f_{p\uparrow}^\dagger f_{q\downarrow}: S^- -
:f_{q\downarrow}^\dagger f_{p\uparrow}: S^+ \right) \nonumber\\
& \ & +\sum\limits_{p,q,r,s} ( \epsilon_p-\epsilon_q+\epsilon_r-\epsilon_s+h(B))  K^\uparrow_{pq,rs}(B)
 \left( :f_{p\uparrow}^\dagger f_{q\downarrow} f_{r\uparrow}^\dagger f_{s\uparrow} : S^- - \text{h.c.} \right)
\nonumber\\
& \ & + \sum\limits_{p,q,r,s} ( \epsilon_p-\epsilon_q+\epsilon_r-\epsilon_s+h(B))  K^\downarrow_{pq,rs}(B) 
 \left( :f_{p\uparrow}^\dagger f_{q\downarrow} f_{r\downarrow}^\dagger f_{s\downarrow} : S^- - \text{h.c.} \right)
\nonumber\\
& \ & +\sum\limits_{p,q,r,s} ( \epsilon_p-\epsilon_q+\epsilon_r-\epsilon_s )  K^\perp_{pq,rs}(B)
 :f_{p\uparrow}^\dagger f_{q\downarrow} f_{r\downarrow}^\dagger f_{s\uparrow} : S^z\ .
\nonumber
\end{eqnarray}
The resulting 2-loop flow equations are worked out in Appendix \ref{Trafo_H}.
The Hamiltonian is diagonalized in a controlled expansion if $\text{max}(|h|,V,T)\gg T_K$, which we assume in the following.
Otherwise the running coupling becomes of ${\cal O}(1)$ and an expansion in its powers is uncontrolled.
\end{widetext}

\subsection{1-loop Scaling Analysis}\label{Scaling_sec_1loop}
The complete set of flow equations cannot be solved analytically due to the complicated momentum dependence.
However, qualitative results for the low energy properties of the system can be worked out analytically.
In the following we derive a simplified scaling picture using the so-called diagonal parametrization\cite{kehrein:056602,kehreinbook,kehrein2007}:
\begin{eqnarray}
J^\perp_{pq}(B) & = & g^\perp_{\overline{pq}}(B) e^{-B(\epsilon_p-\epsilon_q+h(B))^2}\label{diag_param}\\ 
J^{\uparrow/\downarrow}_{pq}(B) & = & g^{\uparrow/\downarrow}_{\overline{pq}}(B) e^{-B(\epsilon_p-\epsilon_q)^2}\nonumber
\end{eqnarray}
where $\overline{pq}=(\epsilon_p+\epsilon_q)/2$. The energy diagonal equations are easily obtained by setting $\epsilon_q=\epsilon_p$ for the
$g^{\uparrow/\downarrow}$ terms and $\epsilon_q=\epsilon_p+h$ for the $g^\perp$ terms.
The diagonal parametrization can be seen as a generalization of the conventional IR-parametrization of scaling theory that allows
for an additional dependence on the energy scale.
Note that the flow of the running coupling does not depend on the momentum index $p$ but on the energy scale $\epsilon_p$: We use
$g_p$ as shorthand notation for $g_{\epsilon_p}$. 

In the following we discuss qualitatively the flow of the 1-loop equations.
We find
\begin{eqnarray}
\frac{dh}{dB} & = & \frac{1}{2} \sum\limits_{p,q} ( n_f(p) + n_f(q)-2n_f(p)n_f(q) )  \\
& \ & \ \ \ \times (\epsilon_p-\epsilon_q+h) (g^\perp_{\overline{pq}})^2  e^{-2B(\epsilon_p-\epsilon_q+h)^2} \nonumber
\end{eqnarray}
for the flow of the magnetic field. Its small shift will be analyzed in the next section.

For a nonzero magnetic field the Kondo couplings will not remain isotropic under the flow. 
The running coupling for parallel scattering is given by
\begin{eqnarray}
\frac{dg_p^\uparrow}{dB} & = & -\sum\limits_r (1-2n_f(r)) (\epsilon_p-\epsilon_r+h) \label{g_up_flow} \\
& \ & \ \ \ \times (g^\perp_{\overline{pr}})^2 e^{-2B(\epsilon_p-\epsilon_r+h)^2}\ , \nonumber
\end{eqnarray}
\begin{eqnarray}
\frac{dg_p^\downarrow}{dB} & = & -\sum\limits_r (1-2n_f(r)) (\epsilon_p-\epsilon_r-h) \\
& \ & \ \ \ \times (g^\perp_{\overline{pr}})^2 e^{-2B(\epsilon_p-\epsilon_r-h)^2}\ , \nonumber
\end{eqnarray}
and for spin-flip scattering one finds
\begin{eqnarray}
\frac{dg_p^\perp}{dB} & = & -\frac{1}{2}\sum\limits_r (1-2n_f(r)) \left(\epsilon_p-\epsilon_r-\frac{h}{2}\right) \label{g_perp_flow}\\
& \ & \ \ \ \times g^\perp_{\overline{\epsilon_r(\epsilon_p+h/2)}}\ g^\uparrow_{\overline{(\epsilon_p-h/2)\epsilon_r}}\ 
e^{-2B(\epsilon_p-\epsilon_r-h/2)^2} \nonumber \\
& \ & -\frac{1}{2}\sum\limits_r (1-2n_f(r)) \left(\epsilon_p-\epsilon_r+\frac{h}{2}\right) \nonumber\\
& \ & \ \ \ \times g^\perp_{\overline{\epsilon_r(\epsilon_p-h/2)}}\ g^\downarrow_{\overline{(\epsilon_p+h/2)\epsilon_r}}\ 
e^{-2B(\epsilon_p-\epsilon_r+h/2)^2} \nonumber\ .
\end{eqnarray}
For convenience we generally drop the $B$-argument of the running coupling and the magnetic field.
At zero temperature the flow of the running coupling can with very good accuracy be simplified using 
$f(x)\exp(-2B(x-c)^2)\approx f(c)\exp(-2B(x-c)^2)$. This approximation removes the $\epsilon_r$ dependence of the running coupling
and the summations in (\ref{g_up_flow})-(\ref{g_perp_flow}) can then be performed and lead to the following expression ($\rho=1$ sets the energy scale):
\begin{eqnarray}
&&\int\limits_{-\infty}^\infty d\epsilon\; (1-2n_f(\epsilon)) (\epsilon+c) e^{-2B(\epsilon+c)^2} \ =  \nonumber\\
&& \ \ =\frac{1}{2B} \left( \frac{e^{-2B(c-V/2)^2}}{1+R}  + \frac{e^{-2B(c+V/2)^2}}{1+1/R}  \right)\ .\label{int_1_loop}\ 
\end{eqnarray}
This yields for parallel scattering
\begin{eqnarray}
\frac{dg^\uparrow_p}{dB} & = & \frac{\left(g^\perp_{\epsilon_p+h/2}\right)^2}{2B} \left( \frac{e^{-2B(\epsilon_p+h+V/2)^2}}{1+R} \right. \nonumber\\
& \ & \ \ \ \left. + \frac{e^{-2B(\epsilon_p+h-V/2)^2}}{1+1/R}  \right)\ ,\label{g_up_diag}
\end{eqnarray}
\begin{eqnarray}
\frac{dg^\downarrow_p}{dB} & = & \frac{\left(g^\perp_{\epsilon_p-h/2}\right)^2}{2B} \left( \frac{e^{-2B(\epsilon_p-h+V/2)^2}}{1+R} \right. \nonumber\\
& \ & \ \ \ \left. + \frac{e^{-2B(\epsilon_p-h-V/2)^2}}{1+1/R}  \right)\ ,\label{g_down_diag}
\end{eqnarray}
and for spin-flip scattering one finds
\begin{eqnarray}
\frac{dg^\perp_p}{dB} & = & \frac{g^\perp_p g^\uparrow_{\epsilon_p-h/2}}{4B} \left( \frac{e^{-2B(\epsilon_p-h/2+V/2)^2}}{1+R} \right. \nonumber\\
& \ & \ \ \ \left. + \frac{e^{-2B(\epsilon_p-h/2-V/2)^2}}{1+1/R}  \right) \nonumber\\
& \ & + \frac{g^\perp_p g^\downarrow_{\epsilon_p+h/2}}{4B} \left( \frac{e^{-2B(\epsilon_p+h/2+V/2)^2}}{1+R} \right. \nonumber\\
& \ & \ \ \ \left. + \frac{e^{-2B(\epsilon_p+h/2-V/2)^2}}{1+1/R}  \right)\ .\label{g_perp_diag}
\end{eqnarray}
The flow of the running coupling is cut off by an exponential decay unless
$\epsilon_p=-(h\pm V/2)$ for $g^\uparrow_p$, $\epsilon_p=h\pm V/2$ for $g^\downarrow_p$, or
$\epsilon_p=\pm(h\pm V)/2$ for $g^\perp_p$.
As a consequence the running coupling is strongly peaked at these energy scales.
In the limit $h=0$ this just corresponds to the strong-coupling behavior of the running coupling at the left and right Fermi level. 
The terms in 2-loop order cut off this strong-coupling behavior as we will see in
the following sections.
Replacing the exponentials in Eqs.~(\ref{g_up_diag})-(\ref{g_perp_diag}) by $\Theta$-step-functions, these equations
become equivalent to the perturbative RG equations derived by Rosch~{\it et al.}\cite{PhysRevLett.90.076804,Rosch2005}.
The different momentum dependence of the running
coupling only leads to subleading corrections.\\

At nonzero temperature $(T>0,V=0)$ one can unfortunately not give a closed expression for
\begin{equation}
\int\limits_{-\infty}^\infty d\epsilon\; \tanh\left(\frac{\epsilon}{2T}\right) (\epsilon+c)e^{-2B(\epsilon+c)^2}\ .
\end{equation}
We therefore only discuss the asymptotic result for $T\gg |h|$. Since we are mainly interested in small energy scales $\epsilon_p\to 0$,
we study the running coupling at the Fermi level only: $g=g^{\perp/\uparrow/\downarrow}_{\epsilon_p=0}$.
For $B\ll T^{-2}$ the terms at high energies $\epsilon\gg T$ give the main contribution to the integral and we obtain the usual zero temperature
scaling equation\cite{0022-3719-3-12-008}
\begin{equation}
\frac{dg}{dB} = \frac{g^2}{2B}\ .
\end{equation}
Note that $\Lambda_{\text{feq}}=B^{-1/2}$.
For $B\gg T^{-2}$ only energies $\epsilon\ll T$ contribute to the integral, since higher energies are cut off by the exponential.
Therefore we linearize the $\tanh$-function and obtain
\begin{equation}
\frac{dg}{dB} = \frac{g^2}{B} \frac{\sqrt{2\pi}}{16} \frac{1}{T\sqrt{B}}\ .
\end{equation}
This implies that the flow of the running coupling effectively stops for $B\gg T^{-2}$ ($T\sqrt{B}\gg 1$). 

To obtain the numerical results shown later we have solved the 2-loop flow equations in diagonal parametrization (\ref{diag_param})
since the solution of the full equations
is very resource intensive. We have verified in selected examples that this approximation agrees extremely well with the full set of equations.

\subsection{2-loop Scaling Analysis I}\label{Scaling_sec_2_loop}
For the case of zero voltage bias and zero temperature, the running couplings $g^{\uparrow (\downarrow)}_p $ are
strongly peaked at $\epsilon_p=-(+) h$, and $g^\perp_p$ at $\epsilon_p=\pm h/2$.
For a qualitative analysis it is sufficient to
replace the momentum dependent
couplings by their peak values. For $B\ll h^{-2}$ we find the well-known 2-loop scaling equations
for the anisotropic Kondo model\cite{0305-4608-4-1-009}
\begin{eqnarray}
\frac{dg_\|(B)}{dB} & = & \frac{g_\perp^2(B)}{2B} - \frac{g_\perp^2(B) g_\|(B)}{4B} \label{B_ll_h}\\
\frac{dg_\perp(B)}{dB} & = & \frac{g_\perp(B) g_\|(B)}{2B} - \frac{g_\perp(B)(g_\|^2(B)+g_\perp^2(B))}{8B}\ , \nonumber
\end{eqnarray}
where $g_\|(B)=g^\uparrow_{-h}(B) = g^\downarrow_{h}(B) $ and $ g_\perp(B)=g^\perp_{\pm h/2}(B) $.
The flow parameter and the remaining effective bandwidth are related by $\Lambda_{\text{feq}}=B^{-1/2}$.
The solution of Eqs.~(\ref{B_ll_h}) is given by $g(B)=g_{\|/\perp}(B)=1/\ln(1/(\sqrt{B}T_K))$ for an initially
isotropic Kondo model.
Additionally, we find a small shift of the magnetic field
\begin{equation}
\frac{dh(B)}{dB} =  -\frac{g_\perp^2(B)}{16 B^{3/2}} \sqrt{2\pi} \text{erf}\left( \sqrt{2B} h(B) \right)\ . \label{h_flow_full}
\end{equation}
For small arguments ($B\ll h^{-2}$) the error function is linear:
\begin{equation}\label{h_renorm}
\frac{dh(B)}{dB} = -\frac{h(B)}{2} \frac{g_\perp^2(B)}{2B}.
\end{equation}
Using $dg_\|/dB=g_\perp^2/(2B)$ Eq.~(\ref{h_renorm}) yields:
\begin{equation}\label{h_renorm_bethe}
h(B) = h_0 \exp\left(-\frac{1}{2} \left( \frac{1}{\ln(1/(\sqrt{B}T_K))} - \frac{1}{\ln(D/T_K)} \right) \right)\ ,
\end{equation}
where $h_0=h(B=0)$.
For large flow parameters $B\gg h^{-2}$ the error function is equivalent to the sign function. This yields a negligible
additional shift of the magnetic field
\begin{equation}
\frac{dh(B)}{dB} =  -\frac{g_\perp^2(B)}{16 B^{3/2}} \sqrt{2\pi}\ .
\end{equation}
of ${\cal O}(h_0/(\ln(h_0/T_K))^2)$.
We can therefore use a constant magnetic field $h(B)=h^*$ 
for $B\gg h^{-2}$, which is determined from Eq.~(\ref{h_renorm_bethe}) by $h^*=h(B=h_0^{-2})$. 
Notice that Bethe Ansatz calculations\cite{PhysRevLett.85.1722} find a shift of the magnetic field to
$h^*\approx h_0(1-1/(2\ln(|h_0|/T_K))$, which is consistent with our result (\ref{h_renorm_bethe}) 
in the scaling limit $D/T_K\rightarrow\infty$ and confirms our approach.

For $B\gg h^{-2}$
we are then left with the flow equations for the coupling constants
\begin{eqnarray}
\frac{dg_\|(B)}{dB} & = & -g_\perp^2(B) g_\|(B) \frac{\sqrt{2\pi}|h^*|}{8\sqrt{B}} \label{B_gg_h}\\
\frac{dg_\perp(B)}{dB} & = & -g_\perp^3(B)  \frac{\sqrt{2\pi}|h^*|}{16\sqrt{B}} \nonumber\ ,
\end{eqnarray}
where we have neglected the 1-loop terms since they only contribute in ${\cal O}(g^2/B)$.
The initial values are given by 
\begin{equation}
g_{\|/\perp}(B_0=(h^*)^{-2})=g^{*}=\frac{1}{\ln(h^{*}/T_{K})}\ .
\end{equation}
The differential equations (\ref{B_gg_h}) are solved by
\begin{eqnarray}
\label{decay28}
g_\|(B) & = & \frac{g^* }{1+\Gamma_{\|}( \sqrt{B}-\sqrt{B_0} ) } \\
g_\perp(B) & = &  \frac{g^*}{\sqrt {1+\Gamma_{\perp}( \sqrt{B}-\sqrt{B_0} ) } }\ ,\nonumber
\end{eqnarray}
with 
\begin{equation}\label{Gamma_h}
\Gamma_{\|}=\Gamma_{\perp} = \frac{\sqrt{2\pi}}{4} (g^*)^2 |h^*| \ .
\end{equation}
In Sect.~\ref{Dot_spin_sec} we will see that up to a prefactor $\Gamma_{\|}$ can be identified with the 
longitudinal and $ \Gamma_{\perp}$ with the transverse spin relaxation rate:
\begin{equation}
\frac{1}{T_{1}}\propto \Gamma_{\|} \ ,\quad \frac{1}{T_{2}}\propto \Gamma_{\perp}
\label{identTGamma}
\end{equation}
We will now already take this identification for granted so that we can compare our result (\ref{Gamma_h})
with literature values. The spin relaxation rate in the limit that the thermal energy is much smaller than
the magnetic energy was first calculated in \cite{Goetze1971} using unrenormalized perturbation theory:
it agrees with (\ref{Gamma_h}) if one uses the same approximation $g^{*}\rightarrow g$. 
Ref.~\cite{Goetze1971} also derived $T_{1}=T_{2}/2$ in this limit, which in our calculation is
hidden in the observation that the proportionality factors in (\ref{identTGamma}) differ due to
the different decay laws in (\ref{decay28}). We will not analyze this in more detail here since
we will later in Sect.~III even calculate the full line shape of the dynamical spin susceptibility. 

\subsection{2-loop Scaling Analysis II}\label{Scaling_sec2}
So far we could use the peaks of the running coupling to derive a simple scaling picture in equilibrium.
Applying a dc-voltage bias yields a splitting of these peaks by $\pm V/2$ since the resonances are pinned 
to the Fermi levels. 
As shown in our previous calculation\cite{kehrein:056602,kehreinbook,kehrein2007} this can be taken into account 
on the rhs of the flow equations by averaging over the splitting of the peaks, e.g.
\begin{equation}
g^{\downarrow}(B) = \frac{1}{V} \int\limits_{h-V/2}^{h+V/2} d\epsilon\; g^\downarrow_\epsilon (B)
\ .
\end{equation}
In the previous section we expanded the flow equations for small flow parameter $B\ll h^{-2}$ and for large flow parameter $B\gg h^{-2}$.
If a dc-voltage bias is applied we find four energy scales that determine small and large $B$, namely $|V+h|$, $|V-h|$, $V$ and $|h|$.
So in principle we would have to discuss the flow equations separately in all five regimes of the flow. However, we can restrict the
following discussion to the initial flow and the flow at very large flow parameter $B\gg B_{0}=\Lambda_0^{-2}$, where 
$\Lambda_0={\rm min}\,\{|V+h|,|V-h|,V,|h|\}$. One can numerically verify that the flow in the intermediate regimes
only leads to small corrections.

In equilibrium at nonzero temperature we can again use the peaks of the running coupling to analyze the flow.
Here the relevant energy scales are given by $T$ and $h\coth(h/(2T))$ and we define 
$\Lambda_0={\rm min}\,\{T,h\coth(h/(2T))\}$.

For small flow parameter $B$ (initial flow) we  find the usual scaling equations (\ref{B_ll_h}) and a small shift of the magnetic field.
In the regime $B\gg \Lambda_0^{-2}$ the 1-loop terms are negligible.
We are then left with the flow equations
\begin{eqnarray}
\frac{dg_\perp(B)}{dB} & = & -\frac{g_\|^2(B)g_\perp(B)}{2\sqrt{B}}c_1-\frac{g_\perp^3(B)}{2\sqrt{B}}c_2\label{flow_2_loop_simple}\\
\frac{dg_\|(B)}{dB} & = & -\frac{g_\perp^2(B)g_\|(B)}{\sqrt{B}}c_2\ . \nonumber
\end{eqnarray}
The initial values are $g_{\|/\perp}(B_0)=g_{\|/\perp}^*$
and the constants are given by
\begin{eqnarray}
\label{eqchV}
c_1(h^*,V) & = & \frac{\sqrt{2\pi}}{4} \frac{V}{(1+R)(1+1/R)} \\
c_2(h^*,V) & = & \frac{\sqrt{2\pi}}{8} \frac{|V+h^*|+|V-h^*|+|h^*|(R+1/R)}{(1+R)(1+1/R)} \nonumber
\end{eqnarray}
in non-equilibrium ($V\neq 0$, but zero temperature $T=0$). In equilibrium ($V=0$, $T\neq 0$) we have
\begin{eqnarray}
c_1(h^*,T) & = & \frac{\sqrt{2\pi}}{4} T \\
c_2(h^*,T) & = & \frac{\sqrt{2\pi}}{8} h^*\coth\left(\frac{h^*}{2T}\right)  \nonumber
\end{eqnarray}
where we have defined $h^*=h(B_0)$.
One easily shows that the solution of
\begin{eqnarray}
\frac{dg_\|(B)}{dB} & = & -\frac{g_\|^3(B)}{\sqrt{B}}c_1-\frac{g_\|^2(B)}{\sqrt{B}}\left( \frac{(g_\perp^*)^2}{g_\|^*}c_2-g_\|^*c_1  \right)\ \ \ \ \label{Abel}\\
g_\perp(B) & = & \sqrt{ g_\|^2(B) \frac{c_1}{c_2} + g_\|(B) \left( \frac{(g_\perp^*)^2}{g_\|^*} -g_\|^* \frac{c_1}{c_2} \right)  } \nonumber
\end{eqnarray}
also solves Eqs.~(\ref{flow_2_loop_simple}).
Thus Eqs.~(\ref{Abel}) are an equivalent formulation of Eqs.~(\ref{flow_2_loop_simple}).
The remaining flow equation is an Abel differential equation of the first kind whose general analytic solution is impossible.
Therefore only asymptotic results can be obtained.
Note that the coupling constants asymptotically flow to zero or to a
nontrivial fixed point $g_\|(B)=g_\|^*-c_2(g_\perp^*)^2/(g_\|^*c_1)$
in Eq.~(\ref{Abel}). This nontrivial fixed point can only be reached in the anisotropic Kondo model with initial
values $g_\perp(B=0)<|g_\|(B=0)|$ and $|h|\lesssim V,T$. It is an unphysical artefact of our approximations in the flow equation
calculation, where higher order terms need to be included in this regime. We will not say something anything about this part of the parameter space 
of the anisotropic Kondo model for the remainder of this paper and now return to the isotropic model.

The equilibrium zero temperature behavior has already been analyzed in the previous section. Notice that the
result (\ref{Gamma_h}) for the spin relaxation rate derived there holds generally when the magnetic energy
dominates, that is for large magnetic field $h\gg V,T$. We next look at the opposite limit of vanishing
magnetic field, $h=0$. The Kondo model remains isotropic and longitudinal and transverse
relaxation rates coincide. Eq.~(\ref{Abel}) is solved by
\begin{equation}\label{g_zero_magnet_ref}
g(B) = \frac{g^*}{\sqrt{1+\Gamma (\sqrt{B}-\sqrt{B_0})}},
\end{equation}
where $g(B)=g_\|(B)=g_\perp(B)$ with the spin relaxation rate $\Gamma=4(g^*)^2c_1$. The equilibrium
finite temperature spin relaxation rate therefore shows the expected Korringa-behavior proportional to temperature 
\cite{Korringa}, while the non-equilibrium zero temperature spin relaxation rate is proportional
to the voltage bias (or more accurately according to (\ref{eqchV}) and (\ref{currentI}): proportional to the current across the dot),
which agrees with the perturbative non-equilibrium RG-result \cite{PhysRevB.70.155301}.

For intermediate values of the magnetic field
we find a competition between the quadratic and the cubic terms in the running coupling in Eq.~(\ref{Abel}).
For very small energies ($B\to\infty$) the quadratic term dominates the flow if it is nonzero and the running couplings decay like $g_\|\sim B^{-1/2}$
and $g_\perp\sim B^{-1/4}$.
If the cubic term dominates both running couplings are proportional to $B^{-1/4}$.
One can still analyze this analytically 
for small magnetic field $V,T\gg |h|$: Eqs.~(\ref{flow_2_loop_simple}) are approximately solved by
\begin{equation}
g_{\|/\perp}(B) = \frac{g_{\|/\perp}^*}{\sqrt{ 1+ \Gamma_{\|/\perp} (  \sqrt{B} - \sqrt{B_0}  ) }}\ ,
\end{equation}
since $g_\|^*\approx g_\perp^*$ and $c_1\approx c_2$.
In the high voltage regime the energy scales where the algebraic decay of the couplings sets in are given by
\begin{eqnarray}
\Gamma_{\|}(h^*,V) & = &  \sqrt{2\pi} (g_{\|}^*)^2 \frac{V}{(1+R)(1+R^{-1})} 
\label{Gamma_V_h}\\
\Gamma_{\perp}(h^*,V) & = & \frac{ \sqrt{2\pi}}{2} (g_\perp^*)^2 \nonumber\\
& \ & \times \frac{|V+h^*|+|V-h^*|+|h^*|(R+R^{-1})}{(1+R)(1+R^{-1})}\ .  \nonumber
\end{eqnarray}
Here we have kept the leading $h$-dependence to show that 
as expected only the spin flip coupling sees both the magnetic field and the voltage bias in its relaxation rate.
We find similar behavior in equilibrium at nonzero temperature for $|h|\ll T$: 
\begin{eqnarray}
\Gamma_{\|}(h^*,T) & = &  \sqrt{2\pi} (g_{\|}^*)^2 T 
\label{Gamma_T_h}\\
\Gamma_{\perp}(h^*,T) & = & \frac{ \sqrt{2\pi}}{2} (g_\perp^*)^2 h^* \coth\left( \frac{h^*}{2T}  \right)\ .  \nonumber
\end{eqnarray}
We have now analytically derived the qualitative scaling behavior and the related energy scales of the Kondo model 
as a function of temperature, voltage bias and magnetic field. In particular we have seen how sufficiently large temperature,
voltage bias (more accurately: a sufficiently large current (\ref{currentI})) or magnetic field can make the Kondo model a weak-coupling problem, where the coupling constants decay
to zero and therefore allow for a controlled solution using flow equations. Similar observations have been
made using other renormalization group techniques\cite{PhysRevB.70.155301,Korb2007,mitra:085342}.
In the next chapter we will turn to a completely 
numerical solution of the flow equations in order to obtain quantitative results. 
This is also the only way to analyze the behavior when the magnetic field is of the same order as the voltage bias
or the temperature, which was excluded in the above analytical discussion. 
Still, the analytical results obtained so far
are important because they will serve as our guidelines to understand and interpret the results of the numerical solution.

\section{Results and Discussion}
We restrict the following discussion of numerical results to symmetric coupling to the leads, $R=1$. The
extension to $R\neq 1$ is straightforward and corresponding results can be obtained without further complications. 
We also want to mention that the approximations in our calculation in this paper do not allow us to obtain a more
accurate result for the current than the previously known expression (\ref{currentI}), and therefore we will
not elaborate on the evaluation of the current in this chapter.

\subsection{Spin-Spin Correlation Function}\label{Dot_spin_sec}
Since the interacting ground state becomes trivial in the $B=\infty$ basis, we transform all operators into the diagonal basis before
calculating their expectation values. We make the following ansatz for $S^z$:
\begin{eqnarray}
S^z(B) & = & h^z(B) S^z  +\frac{M(B)}{2} \label{Ansatz_S_z}\\
& \ & + \sum\limits_{p,q} \gamma_{pq}(B) ( :f_{p\uparrow}^\dagger f_{q\downarrow} : S^- + :f_{q\downarrow}^\dagger f_{p\uparrow} : S^+ )\ , \nonumber
\end{eqnarray}
where $h^z(B=0)=1$, $\gamma_{pq}(B=0)=0$ and $M(B=0)=0$.
For the transformation of the spin operator it is sufficient to use only the first order part of the generator (\ref{Generator}),
that is to neglect terms in $O(J^{2})$ in the generator: In Ref.~\cite{kehrein2007} we showed that this approximation
already yields results including their full leading logarithmic corrections. 

The decay of the spin operator into a different structure under the unitary flow is described by the
flow equation for the coefficient~$h^{z}(B)$:
\begin{eqnarray}
\frac{dh^z(B)}{dB} & = & - \sum\limits_{p,q}( n_f(p)+n_f(q)-2n_f(p)n_f(q) ) \label{flow_eq_h_z_ref}\ \ \ \\
& \ & \ \ \ \times (\epsilon_p-\epsilon_q+h) J^\perp_{pq}(B)\gamma_{pq}(B)\ .\nonumber
\end{eqnarray}
For the flow of the newly generated c-number we find
\begin{eqnarray}
\frac{dM(B)}{dB} & = &  \sum\limits_{p,q} (n_f(p)-n_f(q)) (\epsilon_p-\epsilon_q+h) \label{ODE_M}\\
& \ & \ \ \ \times J^\perp_{pq}(B)\gamma_{pq}(B)\ .\nonumber
\end{eqnarray}
For zero magnetic field the relations $J^\perp_{pq}(B)=J^\perp_{qp}(B)$ and $\gamma_{pq}(B)=-\gamma_{qp}(B)$ are fulfilled.
Using these relations one easily shows $M(B)\equiv0$. We will later see that $M$ is just the magnetization and
therefore it makes sense that
$M$ only becomes nonzero during the flow if an external magnetic field is applied.\\
The flow of the newly generated operator structure in the spin operator is given by
\begin{eqnarray}
\frac{d\gamma_{pq}(B)}{dB} & = & \frac{h^z}{2} (\epsilon_p-\epsilon_q+h) J^\perp_{pq}(B) \label{gamma_pq_fe_ref}\\
& \ & + \frac{1}{4} \sum\limits_r (1-2n_f(r)) \left( (\epsilon_r-\epsilon_p) J^\uparrow_{pr}(B) \right.\nonumber\\
& \ & \ \ \ \left. \times \gamma_{rq}(B)+(\epsilon_r- \epsilon_q) J^\downarrow_{rq}(B) \gamma_{pr}(B) \right)\ .\nonumber
\end{eqnarray}
In the sequel we will focus on the longitudinal spin susceptibility. The calculation of the
transverse part follows exactly the same route and 
the transformation laws of $S^{x/y}(B)$ are given in Appendix \ref{XYZ}.

During the initial flow $h^z(B)$ is nearly unchanged: its flow is only of ${\cal O}(g_\|(B=0)-g_\|(B))$.
For simplicity we restrict the following discussion
to equilibrium and zero temperature; the extension to $V,T>0$ is straightforward.
In lowest order the solution of Eq.~(\ref{gamma_pq_fe_ref}) is formally given by
\begin{equation}
\gamma_{pq}(B) = \frac{1}{2} (\epsilon_p-\epsilon_q+h) \int\limits_0^B dB_1 \; h^z(B_1) J_\perp(p,q,B_1)\ .
\end{equation}
With Eq.~(\ref{flow_eq_h_z_ref}) follows (using diagonal parametrization):
\begin{equation}
\frac{dh^z}{dB} \approx -\frac{\sqrt{\pi}}{8} \frac{g_\perp^2 h^z}{\sqrt{B}} h
\ ,
\end{equation}
where $h$ is the external magnetic field.
Notice the similarity to the 2-loop flow equation for $g_\|$ (\ref{B_gg_h}). Therefore the $S^z$-operator begins to decay on the same
energy scale at which the strong coupling divergence of $g_\|$ is cut off. By a similar argument one can show that the decay of the
$S^{x/y}$-operators is related to the flow of $g_\perp$: the decay starts on the same energy scale that cuts off the strong coupling
divergence of $g_\perp$. We conclude that the energy scales $\Gamma_\|$ and $\Gamma_\perp$ determine the decay of the spin 
operators parallel and perpendicular to the external magnetic field.

It is this observation that relates the decoherence rates $\Gamma_\|$ and $\Gamma_\perp$ to the  physical spin relaxation rates:
$1/T_1$ and $1/T_2$ are defined through the broadening of the resonance poles in the
longitudinal and the transverse dynamical spin susceptibilities \cite{PhysRevB.70.155301}.
Now for $B\sim \Gamma_{\|/\perp}^{-2}$ all excitations
with energy transfer much larger than the decoherence rate are integrated out. Since the spin operator has not yet decayed on this $B$-scale,
the broadening of the resonance pole can therefore not be larger than the corresponding energy scale. The algebraic decay
of the spin operator just corresponds to the broadening of the resonance pole. 
Hence (up to a prefactor) $1/T_{1}\sim \Gamma_{\|}$ and $1/T_{2}\sim \Gamma_{\perp}$.
The width of the broadening of the resonance poles is therefore
(up to an uninteresting prefactor) automatically given by the decoherence rates defined in Sects.~\ref{Scaling_sec_2_loop} and \ref{Scaling_sec2}. 
As already discussed there, our results for the longitudinal and the transverse spin relaxation rates agree
with previous results in the literature in their appropriate limits.

The symmetrized spin-spin correlation function is defined as
$C^{a}(t_1,t_2)=\frac{1}{2} \langle  \{ S^{a}(t_1) , S^{a}(t_2) \} \rangle$ and the response function as
$\chi_{a}(t_1,t_2)=-i \Theta(t_1-t_2) \langle [ S^{a}(t_1) , S^{a}(t_2) ] \rangle$.
In equilibrium these expectation values are evaluated with respect to the ground state or thermal state
defined by $H(B=\infty)$.\cite{kehreinbook} In non-equilibrium they have to be evaluated with
respect to the current-carrying steady state. This was discussed in Ref.~\cite{kehrein2007}: By explicitly
following the time evolution of the initial state until the steady state builds up, we could show that we can
work with the same state as in equilibrium in the present order of our calculation (notice that this does
not hold for the evaluation of the current operator itself). 
\begin{widetext}
The Fourier transformed $S^z$-$S^{z}$ correlation function is therefore both in equilibrium and
non-equilibrium given by 
\begin{eqnarray}
C^{z}(\omega) & = & \frac{\pi(1-\text{sgn}( \tilde{h}))}{2}  \sum\limits_p 
 \left(  \tilde{\gamma}_{\epsilon_p,\epsilon_p+\omega+\tilde{h}}^2 n_f(\epsilon_p)(1-n_f(\epsilon_p+\omega+\tilde{h}))
 +  \tilde{\gamma}_{\epsilon_p,\epsilon_p-\omega+\tilde{h}}^2 n_f(\epsilon_p)(1-n_f(\epsilon_p-\omega+\tilde{h}))\right)\label{corr_func}\\
& \ &  +\frac{\pi(1+\text{sgn}( \tilde{h})}{2} \sum\limits_p 
  \left(  \tilde{\gamma}_{\epsilon_p,\epsilon_p+\omega+\tilde{h}}^2 n_f(\epsilon_p+\omega+\tilde{h})(1-n_f(\epsilon_p)) 
  +  \tilde{\gamma}_{\epsilon_p,\epsilon_p-\omega+\tilde{h}}^2 n_f(\epsilon_p-\omega+\tilde{h})(1-n_f(\epsilon_p)) \right) \nonumber\\
& \ & +\frac{\pi}{2} \tilde{M}^2 \delta(\omega)\ ,\nonumber
\end{eqnarray}
where the tilde denotes the value at $B=\infty$. The corresponding imaginary part of the Fourier transformed
response function is
\begin{eqnarray}
\chi_{z}^{\prime\prime}(\omega) & = & \frac{\pi(1-\text{sgn}( \tilde{h}))}{2}\sum\limits_p
\left(  \tilde{\gamma}_{\epsilon_p,\epsilon_p+\omega+\tilde{h}}^2 
n_f(\epsilon_p)(1-n_f(\epsilon_p+\omega+\tilde{h}))
-  \tilde{\gamma}_{\epsilon_p,\epsilon_p-\omega+\tilde{h}}^2 
n_f(\epsilon_p)(1-n_f(\epsilon_p-\omega+\tilde{h}))\right) \ \ \ \ \\
& \ &  +\frac{\pi(1+\text{sgn}( \tilde{h})}{2}\sum\limits_p 
  \left(  \tilde{\gamma}_{\epsilon_p,\epsilon_p+\omega+\tilde{h}} \ \ 
n_f(\epsilon_p+\omega+\tilde{h})(1-n_f(\epsilon_p)) 
-  \tilde{\gamma}_{\epsilon_p,\epsilon_p-\omega+\tilde{h}}^2 n_f(\epsilon_p-\omega+\tilde{h})
(1-n_f(\epsilon_p)) \right)\ , \nonumber
\end{eqnarray}
\end{widetext}
the real part is accessible via a Kramers-Kronig transformation.
The correlation function is a symmetric function of $\omega$, the imaginary part of the response function is antisymmetric.
Both functions do not depend on the sign of $h$.
In equilibrium the fluctuation dissipation theorem\cite{PhysRev.83.34} relates the imaginary part of the response function and the spin-spin correlation function by
$\chi_{z}^{\prime\prime}(\omega)=\tanh(\omega/(2T)) C^{z}(\omega)$. In non-equilibrium the fluctuation dissipation theorem is violated in general.
For completeness the corresponding expressions for $S^{x/y}$ are given in Appendix \ref{XYZ}.

\begin{figure}
\includegraphics{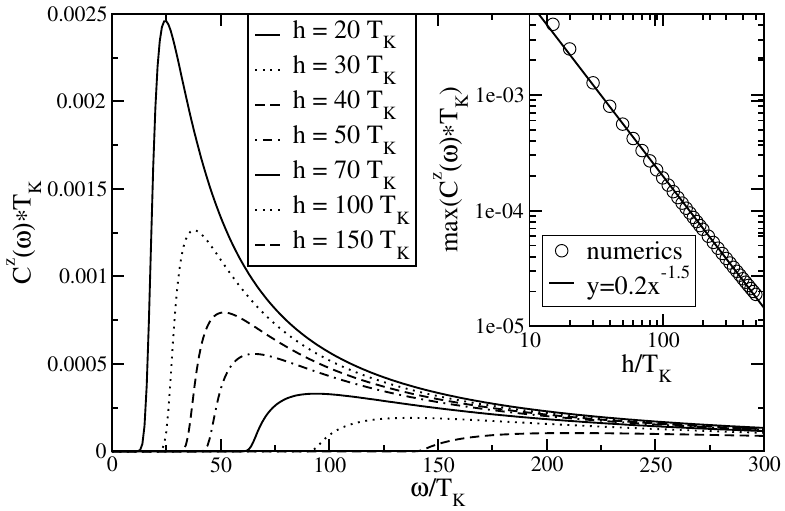}
\caption{Equilibrium $S^z$-$S^{z}$ correlation function for various magnetic fields, V=T=0. The inset shows the 
decay of the peak height, which approximately follows a power law.}\label{C_z_h_fig}
\end{figure}

\begin{figure}
\includegraphics{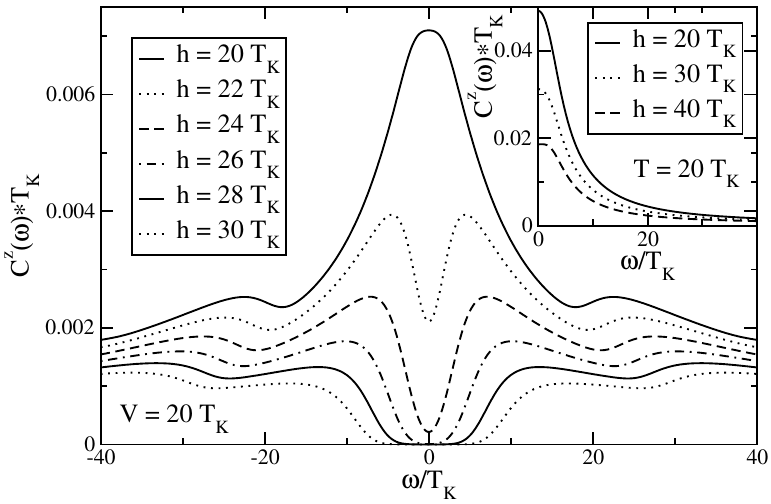}
\caption{Emergence of the large magnetic field behavior of the spin-spin correlation function depicted
in Fig.~\ref{C_z_h_fig} for a non-equilibrium situation ($V=20 T_{K}, T=0$). 
The inset shows typical equilibrium correlation functions for nonzero temperature $T$. Here all features
except the zero frequency peak are smeared out.}\label{corr_h_V_2_fig}
\end{figure}

Typical equilibrium zero temperature spin-spin correlation functions are shown in Fig.~\ref{C_z_h_fig}.
At zero frequency we find a $\delta$-peak with strength $M^2(B=\infty)\pi/2$ in the correlation function (\ref{corr_func})
due to the nonzero spin expectation value
(it is not plotted for obvious reasons). For convenience we assume $h>0$ in the following discussion.
The maximum of the spin-spin correlation function (ignoring the $\delta$-peak at $\omega=0$) is as
expected at $\omega\approx h^{*}$ and it decays with increasing magnetic field (see the inset of
Fig.~\ref{C_z_h_fig}).
For $|\omega| < h^*$ the correlation function vanishes exactly in the present order of the calculation, 
for $|\omega|\gg |h|$ we find $C^z(\omega) \sim 1/(|\omega|(\ln(|\omega|/T_K))^2)$.
Fig.~\ref{corr_h_V_2_fig} shows the buildup of this characteristic behavior of the spin-spin correlation function
also for nonzero voltage bias\cite{kehrein2007} upon increasing the magnetic field. 
The inset shows the corresponding plot in equilibrium for nonzero temperature.
In non-equilibrium we find pronounced peaks
at $|\omega|\sim |h-V|$, $|\omega|\sim |h+V|$, and $|\omega|\sim h$ for $h>V$. The peaks at $\omega\sim \pm |h-V|$ join for
$h\le V$ and build up the zero frequency peak. 

Notice that for nonzero temperature all additional peaks are smeared out (inset of Fig.~\ref{corr_h_V_2_fig}).
This exemplifies a key difference between non-equilibrium and nonzero temperature that will keep
reappearing in other dynamical quantities. The non-equilibrium Fermi function (\ref{defnfp})
retains its characteristic discontinuities, which lead to strong-coupling behavior yielding Kondo-split peaks 
in dynamical quantities. These peaks are only cut off by the decoherence rate and not by voltage or
temperature itself, and therefore much more pronounced. \\

The spin-spin correlation function of the Kondo Model in a magnetic field has so far mainly been studied in the context
of the spin boson model\cite{PhysRevA.41.5383,PhysRevLett.76.1683}.
For high magnetic fields no previous results exist, since high frequencies are difficult to access by
numerical methods like NRG. Paaske~{\it et al.}\cite{PhysRevB.70.155301} studied the transverse dynamical spin susceptibility for high voltage bias,
which can be calculated within our approach using the transformation of the spin operators perpendicular to the magnetic field in Appendix \ref{XYZ}.
Using a Majorana fermion representation Mao~{\it et al.}\cite{PhysRevLett.91.207203} obtained the low 
frequency properties of this correlation function
in the case of dc-voltage bias and nonzero temperature in agreement with our results. 

\subsection{Magnetization}\label{Magnet_sec}
\begin{figure}
\includegraphics{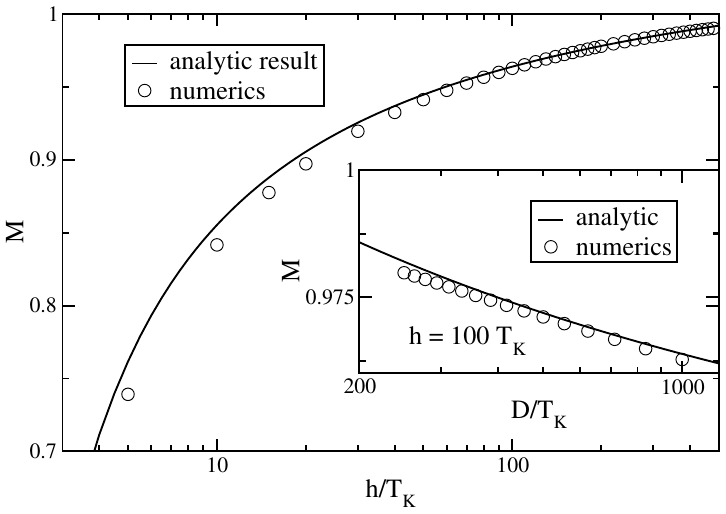}
\caption{The magnetization as a function of the external magnetic field 
calculated from the analytic result (\ref{magnetization_BA}) and from the numerical solution
of the flow equations for $D=10^3 T_K$.
The inset shows the bandwidth dependence of the magnetization at $h=100 T_K$.}\label{h_magn_fig}
\end{figure}

From the ansatz (\ref{Ansatz_S_z}) follows directly that the magnetization of the dot spin is given by $M(B=\infty)$.
However, it turns out that the $S^z$ operator decays slowly with $B$ and therefore also
the magnetization converges slowly, making the analysis difficult. 
Still, we can use the following trick to obtain the leading behavior of the magnetization for $T=V=0$ analytically.
Clearly
\begin{eqnarray}
2\langle S^z\rangle & = & 2 h^z(\infty) \langle \infty | S^z | \infty \rangle +M(\infty) \nonumber\\
& = & h^z(\infty) \mbox{sgn}(h(\infty)) + M(\infty)\ ,\label{Mag_infty}
\end{eqnarray}
where $| \infty \rangle$ is the ground state of $H_0(B=\infty)$. Note that $h^z(\infty)=0$.
For convenience we assume $h>0$ in the following. We rewrite Eq.~(\ref{Mag_infty}) to the form
\begin{eqnarray}
2\langle S^z\rangle & = & h^z(0) + M(0) + \int\limits_0^\infty dB\; \frac{d(h^z(B)+M(B))}{dB}\nonumber\\
& = & 1 - 2 \int\limits_0^\infty dB\; 
\sum\limits_{p,q} n_f(q)(1-n_f(p))\nonumber\\
& \ & \;\; \times (\epsilon_p-\epsilon_q+h) J^\perp_{pq}(B)\gamma_{pq}(B)\ .
\end{eqnarray}
Using the parametrization
\begin{eqnarray}
J^\perp_{pq}(B) & \approx & g_\perp(B) e^{-B(\epsilon_p-\epsilon_q+h)^2} \label{Param_gamma}\\
\gamma_{pq}(B) & \approx & \frac{g_\perp(B) }{2(\epsilon_p-\epsilon_q+h)}
\left( 1-e^{-B(\epsilon_p-\epsilon_q+h)^2} \right) \nonumber
\end{eqnarray}
we find
\begin{equation}
2 \langle S^z \rangle \approx 1 -  \int\limits_{D^{-2}}^\infty dB\;  \frac{g_\perp^2(B)}{4B} f(B)\ ,
\end{equation}
where $f(B)=1$ for $B\ll h^{-2}$ and $f(B)=0$ for $B \gg h^{-2}$.
Neglecting higher order corrections we find
\begin{eqnarray}
2\langle S^z \rangle & \approx & 1 - \int\limits_{D^{-2}}^{h^{-2}} dB\; \frac{g_\perp^2(B)}{4B} \ .
\end{eqnarray}
With $ dg_\|(B)/dB=g_\perp^2(B)/(2B) $ follows
\begin{equation}\label{magnetization_BA}
2 \langle S^z \rangle \approx  1 - \frac{1}{2 \ln(h/T_K)} +\frac{1}{2 \ln(D/T_K)}\ ,
\end{equation}
which (for $D/T_K\to\infty$) is to leading order the Bethe Ansatz result\cite{PhysRevLett.46.356}.
Fig.~\ref{h_magn_fig} shows the good agreement between the analytic expression and numerical results for high magnetic fields.
For fields of ${\cal O}(10\;T_K)$ we see deviations from the analytic result due to the perturbative nature of our approach.
The inset shows the bandwidth dependence of the magnetization in good agreement with Eq.~(\ref{magnetization_BA}).\\
Unfortunately, a similar analytical calculation for $T,V>0$ has not been possible.
We will present numerical results in Sect.~\ref{magnet_numerics_ref}.

\subsection{T-Matrix}\label{T-Matrix_sec}
The scattering of conduction band electrons from lead $\alpha$ to lead $\beta$ is described by the T-matrix
${\cal T}_{\alpha\beta,\sigma}(\omega)$. It is defined via the electron Greens function
\begin{equation}
{\cal G}_{\alpha\beta,\sigma}(\omega) = {\cal G}_{\alpha,\sigma}^{(0)}(\omega) \delta_{\alpha,\beta} +
{\cal G}_{\alpha,\sigma}^{(0)}(\omega) {\cal T}_{\alpha\beta,\sigma}(\omega) {\cal G}_{\beta,\sigma}^{(0)}(\omega)
\ .
\end{equation}
If the Hamiltonian (\ref{H_ini}) is derived from an Anderson impurity model only one eigenvalue
of the T-matrix is nonzero\cite{PhysRevB.70.155301}. 
The imaginary part of the T-matrix is given by\cite{PhysRevLett.85.1504}
\begin{eqnarray}\label{T_Matrix_equ}
\text{Im}[T_\sigma(\omega)]  =  - \int\limits_{-\infty}^{\infty} dt\; \Theta(t) \langle \{ O_\sigma(t) , O_\sigma^\dagger(0)
\} \rangle e^{i\omega t}\ ,
\end{eqnarray}
where
\begin{eqnarray}
\label{defO}
O_\uparrow(B) & = & \sum\limits_k ( U^\perp_k(B) f_{k\downarrow} S^- + U^\uparrow_k(B) f_{k\uparrow} S^z )\\
O_\downarrow(B) & = & \sum\limits_k ( V^\perp_k(B) f_{k\uparrow} S^+ - V^\uparrow_k(B) f_{k\downarrow} S^z )\ , \nonumber
\end{eqnarray}
$U_k^{\perp/\uparrow}(B=0)=J^{\perp/\uparrow}/2$ and
$V_k^{\perp/\downarrow}(B=0)=J^{\perp/\downarrow}/2$.
In lowest order the flow equations for the spin up component are given by
\begin{equation}
\frac{dU_p^\uparrow}{dB} = -\frac{1}{2} \sum\limits_{r} (1-2n_f(r))(\epsilon_p-\epsilon_r+h) U_p^\perp J_{rp}^\perp\ ,
\end{equation}
\begin{eqnarray}
\frac{dU_p^\perp}{dB} & = & -\frac{1}{4} \sum\limits_{r} (1-2n_f(r)) (\epsilon_p-\epsilon_r+h) U_r^\uparrow J_{kp}^\perp\nonumber\\
& \ & -\frac{1}{4} \sum\limits_{r} (1-2n_f(r)) (\epsilon_p-\epsilon_r) U_p^\perp J_{rp}^\downarrow\ .
\end{eqnarray}
\begin{figure}
\includegraphics{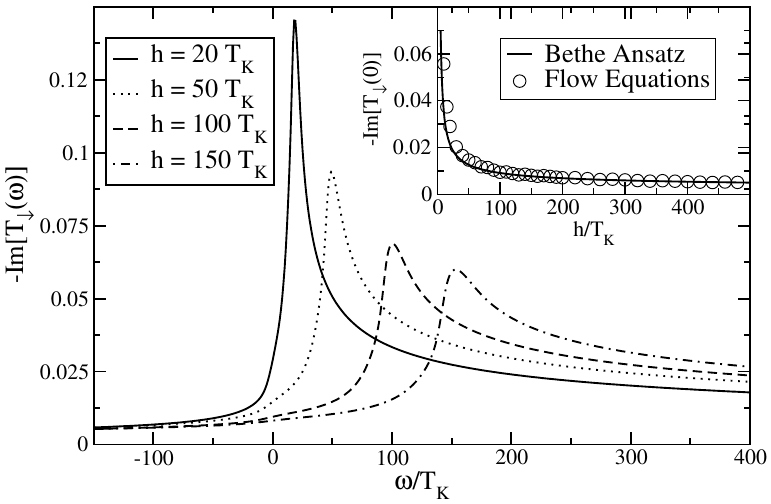}
\caption{Imaginary part of the spin-down T-matrix as a function of the magnetic field ($V=T=0, D = 10^3 T_K$).
The inset shows a comparison with the Bethe Ansatz result
for $\omega=0$.}\label{t_m_h_fig}
\end{figure}
Comparing the latter equations with Eqs.~(\ref{g_up_flow}) and (\ref{g_perp_flow}) one already notices their similarity to the flow of the
running coupling in 1-loop order. In the following we work out the details. Using the approximations from Sect.~\ref{Scaling_sec_1loop}
we find
\begin{eqnarray}
\frac{dU_p^\uparrow}{dB} & = & \frac{g_{\epsilon_p+h/2}^\perp U_{\epsilon_p+h}^\perp}{2B} \left( \frac{e^{-B(\epsilon_p+h+V/2)^2}}{1+R} \right. \nonumber\\
& \ & \ \ \ \left. +  \frac{e^{-B(\epsilon_p+h-V/2)^2}}{1+1/R} \right)\ .
\end{eqnarray}
Neglecting a factor two in the exponential, this equation is equivalent to Eq.~(\ref{g_up_diag}) provided
$U_p^\uparrow=g_p^\uparrow/2$ and $U_{\epsilon_p+h/2}^\perp=g_p^\perp/2$. Analyzing the flow of the spin-flip component we find
\begin{eqnarray}
\frac{dU_{\epsilon_p+h/2}^\perp}{dB} & = & \frac{g_{p}^\perp U_{\epsilon_p-h/2}^\uparrow}{4B} \left( \frac{e^{-B(\epsilon_p-h/2+V/2)^2}}{1+R} \right. \nonumber\\
& \ & \ \ \ \left. +  \frac{e^{-B(\epsilon_p-h/2-V/2)^2}}{1+1/R} \right) \nonumber\\
& \ & + \frac{g_{\epsilon_p+h/2}^\downarrow U_{\epsilon_p+h/2}^\perp}{4B} \left( \frac{e^{-B(\epsilon_p+h/2+V/2)^2}}{1+R} \right. \nonumber\\
& \ & \ \ \ \left. +  \frac{e^{-B(\epsilon_p+h/2-V/2)^2}}{1+1/R} \right)\ .
\end{eqnarray}
Again neglecting the factor two in the exponential, this equation is equivalent to the 1-loop flow equation (\ref{g_perp_diag})
for $g_p^\perp$.
One easily shows that higher order terms in the transformation of the operator (\ref{defO})
have the same effect on the flow as the 2-loop terms in the
transformation of the Hamiltonian. The calculation for nonzero temperature is again more difficult, nevertheless we find the same relations
between the flow of the operator and the running coupling.

Doing an analogous argument for the $V$-terms we identify
\begin{eqnarray}
U^\uparrow_p(B) = \frac{g^\uparrow_p(B)}{2} & ,\; & U^\perp_p(B) =  \frac{g^\perp_{\epsilon_p-h/2}(B)}{2}\ , \\
V^\downarrow_p(B) = \frac{g^\downarrow_p(B)}{2} & ,\; & V^\perp_p(B) =  \frac{g^\perp_{\epsilon_p+h/2}(B)}{2}\ . \nonumber
\end{eqnarray}
\begin{figure}
\includegraphics{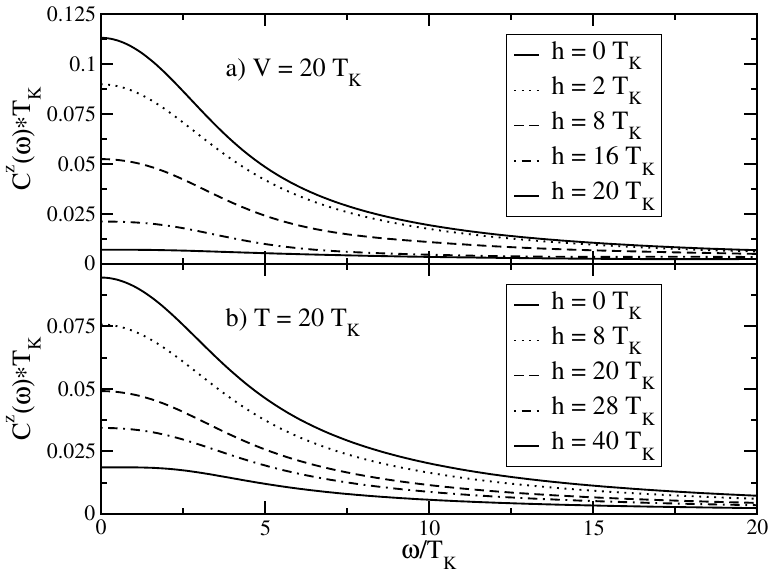}
\caption{Decay of the zero frequency peak in the spin-spin correlation function with increasing magnetic field,
a) $V=20T_K$, b) $T=20T_K$.}\label{corr_h_V_T_fig}
\end{figure}
Therefore the $O_\sigma$ operators completely decay into more complicated objects for $B\to\infty$.
Since calculating the latter is resource intensive (three momentum indices), it is more economic to
evaluate the T-matrix at the decoherence scale\cite{PhysRevB.68.014430}, where the decay of the couplings sets in and
higher order terms in the transformation of the observable are not yet important:
\begin{eqnarray}
&&\text{Im}[T_\sigma(\omega)]  \approx - \frac{\pi}{16}   \left( (\hat{g}^\sigma_\omega)^2 \phantom{\hat{g}^\perp_{\omega+\sigma \hat{h}/2})^2}\right. 
\label{eqImT}\\
&& \ \ \ \ \left. +
2 (\hat{g}^\perp_{\omega+\sigma \hat{h}/2})^2  \left( 1+\sigma 2\langle S^z \rangle (2\hat{n}_f(\omega+\sigma \hat{h})-1 ) \right)\right)\ . \nonumber
\end{eqnarray}
Here the hat denotes functions at the decoherence scale.
Though the further flow leads to a decay of $O_\sigma$, the spectral function remains unchanged for $B>\Gamma_{\|/\perp}^{-2}$, where
$\Gamma_{\|/\perp}$ is the dominant decoherence scale. 
In (\ref{eqImT}) we can replace the expectation value of $S^z$ at the decoherence scale by the magnetization of the system since the $S^z$ operator
decays noticeably only for $B\gg\Gamma_{\|/\perp}^{-2}$.

As suggested by Rosch~{\it et al.}\cite{PhysRevB.68.014430} we use Fermi functions broadened by the decoherence scale $\Gamma_\perp$ to describe
the distribution function for the $f$-operators at the decoherence scale $\hat{n}_f(\omega)$. 
This avoids the costly full numerical solution to $B\rightarrow\infty$ and yields results that
are virtually identical. 
In equilibrium at small temperature $T\ll |h|$ 
the distribution function is then given by $\hat{n}_f(\omega)=f_\Gamma(\omega)$,
where $f_\Gamma(\omega)=1/2-\arctan(\omega/\Gamma_{\perp})/\pi$. At high temperature $T\gg |h|$ the spin expectation value
$\langle S^z \rangle$ vanishes. Then the distribution function only enters in subleading order. 
Note that the imaginary part of the T-matrix in general depends only weakly on the details of the broadening scheme.
In non-equilibrium the step functions at both chemical potentials have to be broadened yielding
$\hat{n}_f(\omega)=f_\Gamma(\omega+V/2)/(1+R)+f_\Gamma(\omega-V/2)/(1+1/R)$ for the distribution function.
The additional factor of $1/4$ in comparison with the result obtained by Rosch~{\it et al.}\cite{PhysRevB.68.014430} 
is due to our different definition of $J$.
For symmetric coupling $R=1$ the spin-up and the spin-down component are related by
$\text{Im}[T_\uparrow(\omega)]=\text{Im}[T_\downarrow(-\omega)]$.

\begin{figure}
\includegraphics{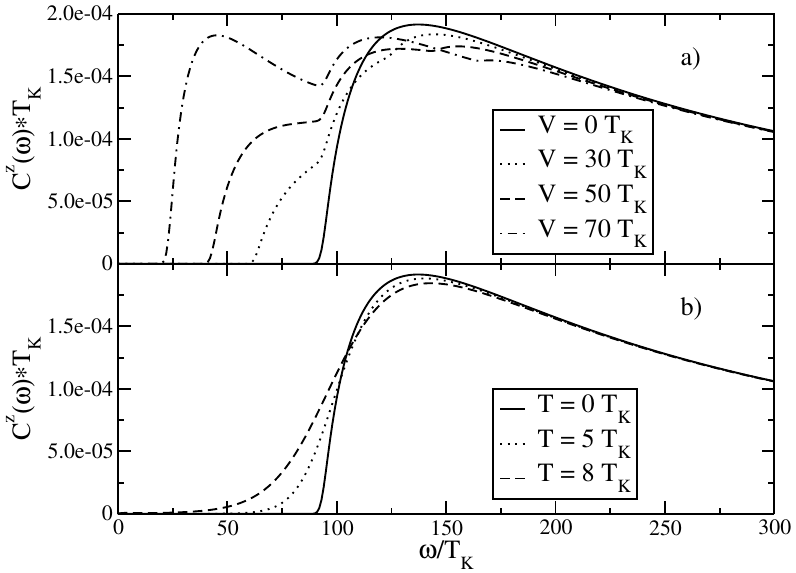}
\caption{a) Kondo splitting of the sharp edge in the spin-spin correlation function (magnetic field $h=100T_K$) due to a small voltage bias.
b) Small temperature only leads to broadening and no Kondo splitting.}\label{corr_h_V_T_2_fig}
\end{figure}

The imaginary part of the T-matrix and the spectral function are related by $A_\sigma(\omega)=-\text{Im}[T_\sigma(\omega+i\delta)]/\pi$.
Fig.~\ref{t_m_h_fig} shows spectral functions for several values of the magnetic field.
They are strongly peaked at $\omega \sim h$.
Rosch~{\it et al.}\cite{PhysRevB.68.014430} studied this structure in detail by analyzing the spectral function 
normalized to $1$ as a function of
$\omega/h$. Since we have included the shift of the magnetic field, we will do likewise as a function of $\omega/h^*$.
In agreement with the results derived by Rosch~{\it et al.}\cite{PhysRevB.68.014430}
we find that the width of the left flank is approximately proportional to the
decoherence rate (\ref{Gamma_h}), leading to a sharpening of the left flank for increasing $h$, while the width of the right flank
increases for increasing $h$.

The imaginary part of the T-matrix at zero frequency is related to the magnetization via the Friedel sum rule.\cite{PhysRev.150.516,Nozieres1974}
Inserting the leading term
of the Bethe Ansatz result\cite{andrei1982} one finds $\text{Im}[\hat{T}_\sigma(0)]=-\sin^2\left( \pi/(4\ln(h/T_K))\right)/\pi$.
The inset of Fig.~\ref{t_m_h_fig} shows a comparison between the Bethe Ansatz and the flow equation result.
Again we find very good agreement for high magnetic fields and deviations for fields of ${\cal O}(10\;T_K)$.

For large frequencies the spectral function decays proportional to $1/(\ln(\omega/T_K))^2$,
\cite{LoganDickens,Dickens2001,PhysRevB.68.014430} which is consistent with our results. 
Also Bethe Ansatz calculations\cite{PhysRevLett.85.1722} show that the maximum of the
spin-down spectral function is at $h^*\approx h(1-1/(2\ln(h/T_K))$,
which is consistent with our shift of the magnetic field (\ref{h_renorm_bethe}) in the scaling limit $D/T_K\to\infty$.

\subsection{Voltage Bias vs. Temperature}\label{V_vs_T_sec}
\begin{figure}
\includegraphics{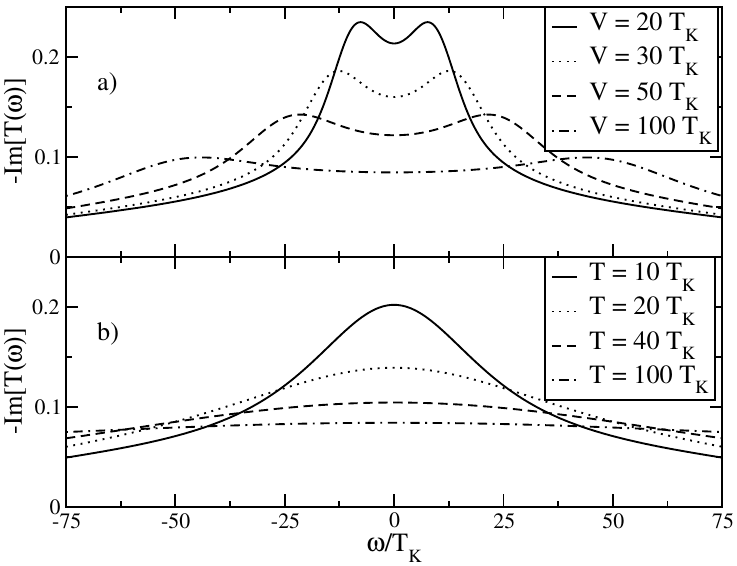}
\caption{Sum of both spin components of the T-matrix at zero magnetic field for various values of a) voltage bias and b) temperature.}\label{t_m_V_T_fig}
\end{figure}
\subsubsection{Spin-Spin Correlation Function}
The spin-spin correlation function at zero magnetic field ($V$ or $T\gg T_K$) shows a zero frequency peak\cite{kehreinbook,kehrein2007}.
In Fig.~\ref{corr_h_V_T_fig}
we show its decay due to an applied magnetic field.
Again the zero frequency $\delta$-peak in Eq.~(\ref{corr_func}) is not plotted.
The sum rule
\begin{equation}\label{sum_rule_ref}
\frac{\pi}{2} = \int\limits_{-\infty}^\infty d\omega\; C^z(\omega) = \tilde{M}^2\frac{\pi}{2} + \int\limits_{-\infty}^\infty d\omega\; C_{\gamma}^z(\omega)
\end{equation}
is not fulfilled exactly since we neglect higher order terms in the transformation of $S^z$. The error is typically of order one percent.
Here $C_{\gamma}^z(\omega)$ denotes the $\tilde{\gamma}_{pq}$ terms in Eq.~(\ref{corr_func}).

For increasing magnetic field the magnetization $\tilde{M}$ increases. Due to the sum rule and the fact that $C_\gamma^z(\omega)$ is a
non-negative function, an increase of $\tilde{M}$ must lead to a decrease of $C_\gamma^z(\omega)$, leading to a decay of the correlation function
for $\omega\neq 0$.

At first glance the decay of the zero frequency peak looks similar for both the equilibrium and the non-equilibrium case.
Only the relative decay of the maximum as a function of $V/h$ and $T/h$ seems to be different.
On closer inspection we find additional peaks at $|\omega|\sim |h|$ for $V>|h|$:
their height increases with the magnetic field, see Fig.~\ref{corr_h_V_2_fig}. In equilibrium for nonzero temperature these peaks are
smeared out. For high frequencies we find the usual $C^z(\omega) \sim 1/(|\omega|(\ln(|\omega|/T_K))^2)$ behavior.

In Fig.~\ref{corr_h_V_T_2_fig} a) we show the Kondo splitting of the sharp edge at $\omega=h^{*}$ in the correlation function 
due to an applied small voltage bias. The two new peaks are
located at $|\omega| \sim |h^{*}\pm V|$. On the other hand, for small temperature we again only find a broadening effect.

\subsubsection{T-Matrix}
\begin{figure}
\includegraphics{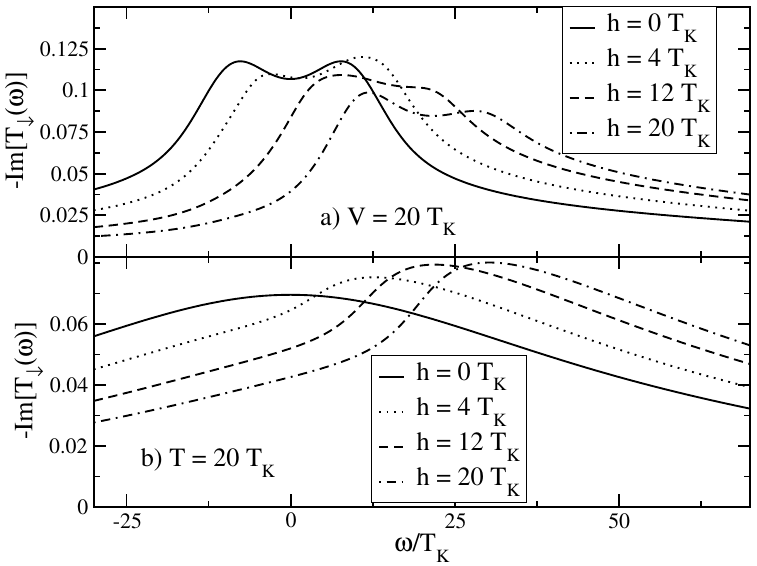}
\caption{a) Magnetic field shift of the Kondo split zero frequency peak of the spectral function for $V=20T_K, T=0$. b) 
Nonzero temperature in equilibrium (here: $T=20T_K, V=0$) leads to broadening.}\label{t_m_h_V_T_fig}
\end{figure}
Fig.~\ref{t_m_V_T_fig} depicts the sum of both spin components of the T-matrix for vanishing external
magnetic field, that is
the full impurity spectral function.
For nonzero voltage bias ($V\gg T_K$) and zero magnetic field one finds the characteristic Kondo split
peaks at $\omega\sim \pm V/2$.
For nonzero temperature one only observes the expected broadening of the zero frequency peak.
These observations are consistent
with the results obtained by NRG\cite{PhysRevLett.85.1504} and perturbative RG\cite{PhysRevB.70.155301}.
Applying a small magnetic field leads to a shift with the
magnetic field strength and an asymmetric deformation of the peaks. Typical curves are shown in Fig.~\ref{t_m_h_V_T_fig}.
\begin{figure}
\includegraphics{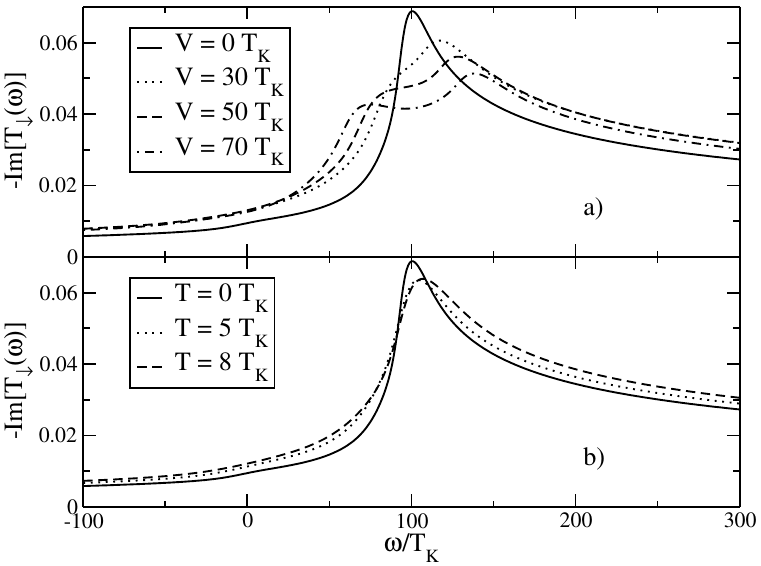}
\caption{a) Splitting of the peak in the spectral function due to a small voltage bias
(external magnetic field $h=100 T_{K}$). b) Again temperature only leads to broadening.}\label{t_m_h_V_T_2_fig}
\end{figure}
For large magnetic fields we have 
already discussed in Sect.~\ref{T-Matrix_sec} how
the spectral function of the equilibrium zero temperature Kondo model develops a pronounced peak at $\omega \sim h$,
see Fig.~\ref{t_m_h_fig}.
In Fig.~\ref{t_m_h_V_T_2_fig} a) we show the splitting of this peak into two peaks at $\omega \sim h\pm V/2$ due to a small voltage bias.
Again, applying a small temperature only leads to a broadening of the peak.\\

We can see that it is straightforward to resolve sharp features in the dynamical quantities at large
frequencies using the flow equation approach, which is notoriously difficult using NRG. For example
the finite temperature broadening in Fig.~\ref{corr_h_V_T_2_fig} and our results for
the T-matrix with magnetic field plus voltage bias or nonzero temperature have not been previously
obtained using other methods.

\subsubsection{Magnetization}\label{magnet_numerics_ref}
In principle the magnetization can be extracted from the spin-spin correlation function via the sum rule (\ref{sum_rule_ref}).
However, due to the approximations in our calculation
the sum rule is not exactly fulfilled and we were only able to extract qualitative results via this route.
More accurate results can be obtained by analyzing the flow of $M(B)$ directly.
We were able to reproduce previously known results from Bethe Ansatz and non-equilibrium perturbation theory.
\begin{figure}
\includegraphics{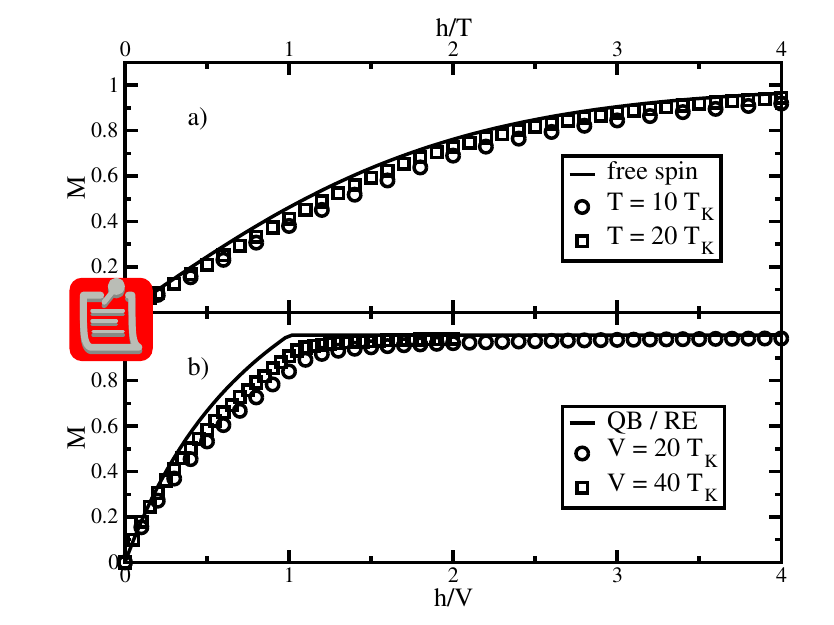}
\caption{The magnetization as a function of the external magnetic field for a) fixed temperature
and b) fixed voltage bias for $D=100T_K$.
For increasing temperature one approaches the free spin behavior as expected. 
For increasing voltage bias one approaches the non-equilibrium perturbation theory
result based on a quantum Boltzmann equation plus rate equations (QB/RE)\cite{paaske:155330},
which becomes exact in the limit $V/T_{K}\rightarrow\infty$.}\label{M_V_T_fig}
\end{figure}

In equilibrium the exact magnetization is accessible by solving the Bethe Ansatz equations\cite{filyov1981,PhysRevLett.49.497,PhysRevLett.46.356}.
Assuming $h>0$ the asymptotic results relevant for this paper
are given by the zero temperature magnetization $M(h,T=0)=1-1/(2\ln(h/T_K))$ for $h\gg T_K$, 
and the high temperature magnetization
$M(h,T)=\tanh(h/(2T))$ for $T\gg T_K$ and $T\gg h$. The high temperature result is of course just 
the magnetization of a free spin.

Previous non-equilibrium perturbation theory calculations\cite{PhysRevB.66.085315,paaske:155330,Reininghaus2009} 
in the limit $V\gg T_K$ or $|h|\gg T_K$  found
$M_{\text{pt}}(h,V)=4h/(2|h|+|h+V|+|h-V|)$ for the magnetization.
Here the important logarithmic corrections at zero voltage bias are missing since $M_{\text{pt}}(h,V)=\text{sgn}(h)$ for $V<|h|$.

\begin{figure}
\includegraphics{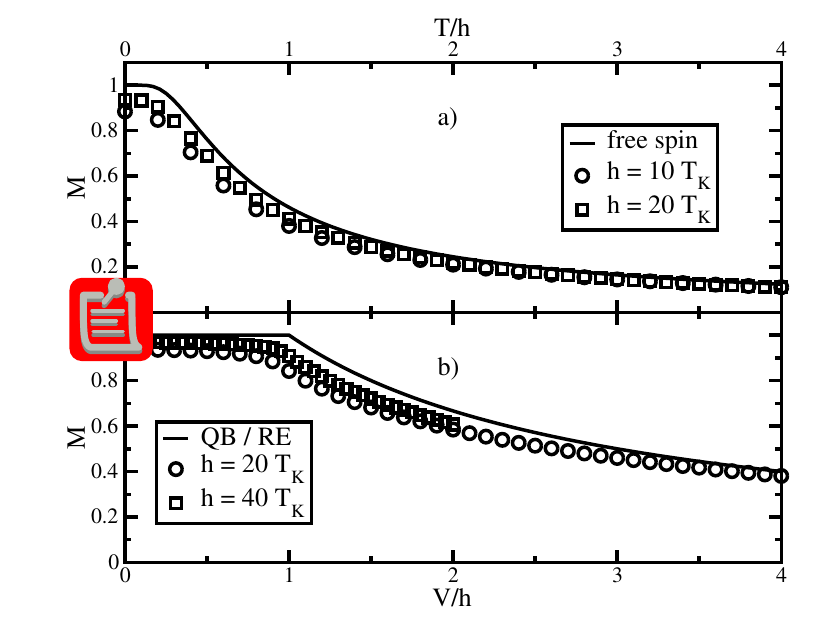}
\caption{Crossover from the equilibrium zero temperature magnetization a) to the high temperature and
b) to the high voltage result for $D=100T_K$.}\label{M_h_V_T_fig}
\end{figure}

In Sect.~\ref{Magnet_sec} we have already derived the zero temperature magnetization within the flow equation framework.
Figs.~\ref{M_V_T_fig} and \ref{M_h_V_T_fig} show the crossover between the equilibrium zero temperature result 
and the asymptotic high temperature result or the asymptotic high voltage bias result.
These crossovers are smooth and show the expected reduction of the magnetization for finite $V<|h|$
that are missing in previous calculations. It should be noted that there is a noticeable dependence of 
the results in Figs.~\ref{M_V_T_fig} and \ref{M_h_V_T_fig} on the bandwidth similar to Fig.~\ref{h_magn_fig}.

\section{Conclusion}
In this paper we have employed the flow equation approach to derive a consistent scaling picture
of the equilibrium and  non-equilibrium Kondo model in its weak-coupling regime. 
The weak-coupling regime is realized for sufficiently large voltage bias~$V$, magnetic field~$h$
or temperature~$T$ as compared to the equilibrium Kondo temperature: $\max(I,|h|,T)\gg T_{K}$, where
$I$ is the current (\ref{currentI}) across the dot.
Our calculation allowed for the evaluation of static and dynamical quantities including their leading logarithmic
corrections. Specifically, we have studied the spin-spin correlation function, the magnetization and
the T-matrix as functions of $V,h$ and $T$ and explored various crossover regimes. 

As emphasized by Millis et al.\cite{mitra:085342,segal:195316}, the non-equilibrium noise
generated by the steady state current across a quantum impurity can to leading order be approximated by thermal noise
with an effective temperature $T_{\rm eff}(I)$, but with important differences between
non-equilibrium noise and thermal noise occuring beyond leading order. We could see
this explicitly in many of our dynamical quantities, where non-equilibrium conditions due to
a voltage bias lead to effects like Kondo splitting and strongly enhanced logarithmic corrections
(this is also especially noticeable in the static spin susceptibility, see Refs.~\cite{kehrein2007,Reininghaus2009}).

As a final comment we want to mention that while the flow equation approach has to rely on
numerical evaluations of complicated sets of differential equations (at least if one is interested in quantitative
results beyond leading order), it does allow one to study all combinations of the parameters
voltage bias, temperature and magnetic field in one framework. As an outlook this should
be useful for investigating other more complex quantum dot structures in the future.

\begin{acknowledgments}
We acknowledge valuable discussions with 
V.~K\"orting, J.~Paaske, A.~Rosch, and H.~Schoeller.
This work was supported through SFB/TR~12 of the Deutsche 
Forschungsgemeinschaft (DFG), the Center for Nanoscience (CeNS)
Munich, and the German Excellence Initiative via the
Nanosystems Initiative Munich (NIM).
\end{acknowledgments}

\appendix

\section{Normal Ordering}\label{Norm_Order}
In this section we briefly sum up some properties of normal ordered operators
that are frequently used in flow equation calculations. For more details
we refer to Ref.~\cite{kehreinbook}

In the following $A_p$ denotes creation and annihilation operators, the $\alpha$'s are c-numbers and $ P( \{ A_p  \} )$
is a product of operators from the set $\{A_p\}$. 
The rules for Wick's normal-ordering are given by:
\begin{enumerate}
\item Numbers are unchanged:
\begin{equation}
: \alpha : = \alpha
\end{equation}
\item Normal-ordering is linear
\begin{eqnarray}
& \ & : \alpha_1 P_1( \{ A_p  \} ) + \alpha_2 P_2 ( \{ A_p  \} ) : \; = \\
& \ & \;\;\;\;\alpha_1 :P_1( \{ A_p  \} ) : +  \alpha_2 : P_2 ( \{ A_p  \} ) : \nonumber
\end{eqnarray}
\item Recurrence relation
\begin{eqnarray}
A_q : P(\{ A_p \}): & = & : A_q P(\{ A_p \}) : \label{Wick_Rec}\\
& \ & \; + \sum\limits_r C_{qr} : \frac{\partial P(\{ A_p \})}{\partial A_r } :\ ,\nonumber
\end{eqnarray}
\end{enumerate}
where
\begin{equation}
C_{qr} = \langle \Psi | A_q A_r | \Psi \rangle
\end{equation}
for a pure reference state $| \Psi \rangle$ or
\begin{equation}
C_{qr} = \text{Tr} (  \rho A_q A_r )
\end{equation}
for some mixed state described by the density matrix $\rho$.
Typically the ground state of the non-interacting Hamiltonian is chosen as reference state $| \Psi \rangle$.\\

From the recurrence relation (\ref{Wick_Rec}) one can derive Wick's first theorem
\begin{eqnarray}
: A_{p_1} \hdots A_{p_n} :  \; = \; \left( A_{p_1} - \sum\limits_{q_1} C_{p_1 q_1}  \frac{\partial}{\partial A_{q_1}} \right) \hdots & \ & \\
\times \left( A_{p_{n-1}} - \sum\limits_{q_{n-1}} C_{p_{n-1} q_{n-1}}  \frac{\partial}{\partial A_{q_{n-1}}} \right) A_{p_n}\ .& \ & \nonumber
\end{eqnarray}
From this relation follows that the commutation of neighboring fermionic operators picks up a minus sign, bosonic operators commute.
The product of two normal-ordered objects can be calculated from Wick's second theorem. The fermionic version is given by
\begin{eqnarray}
:  P_1( \{ A_p  \} ) : : P_2( \{ A_p  \} ) : \; = \; : \exp\left(  \sum\limits_{r,s}  C_{rs} \frac{\partial^2}{\partial B_s  \partial A_r }  \right) & \ & \; \label{NormalOrderProduct}\\
\times P_1( \{ A_p  \} ) P_2( \{ B_p  \} ) :  \Big|_{A=B}\ .  & \ & \nonumber
\end{eqnarray}

\section{Transformation of the Hamiltonian}\label{Trafo_H}
The derivation of the flow equations for the Hamiltonian (\ref{Ansatz}) is straightforward.
Only some preliminary relations are needed.
Products of spin operators are easily calculated using the standard spin operator algebra. The relations
\begin{eqnarray}
\; [ AS^- , B S^z ] & = & \frac{1}{2} \{ A,B \} S^- \label{Spin_Algebra}\\
\; [ AS^+ , B S^z ] & = & -\frac{1}{2} \{ A,B \} S^+ \nonumber\\
\; [ A S^+ , B S^- ] & = & \{ A,B \} S^z + \frac{1}{2} [ A , B ] \nonumber
\end{eqnarray}
are fulfilled for arbitrary (linear) operators $A,B$ that commute with the spin operators.

Using Eq.~(\ref{NormalOrderProduct}) the following relations are easily derived.
For the 1-loop calculation, the commutator
\begin{eqnarray}
[:c_{1^\prime}^{\dagger}c_{1}:,:c_{2^\prime}^{\dagger}c_{2}:] & = & 
:c_{1^\prime}^{\dagger}c_{2}:\delta_{1,2^\prime}
- :c_{2^\prime}^{\dagger}c_{1}:\delta_{1^\prime,2}  \;\;\; \\
& \ & +\delta_{1^\prime,2}\delta_{1,2^\prime}(n(1^\prime)-n(1)) \nonumber
\end{eqnarray}
is needed. Due to the spin operator algebra (\ref{Spin_Algebra}) also the anticommutator has to be calculated:
\begin{eqnarray}
& \ & \{:c_{1^\prime}^{\dagger}c_{1}:,:c_{2^\prime}^{\dagger}c_{2}:\} \; = \;
2:c_{1^\prime}^{\dagger}c_{1}c_{2^\prime}^{\dagger}c_{2}: \\
& \ & \; \; \; + \delta_{1,2^\prime}(1-2n(1)):c_{1^\prime}^{\dagger}c_{2}: \nonumber\\
& \ & \; \; \; + \delta_{2,1^\prime}(1-2n(1^\prime)):c_{2^\prime}^{\dagger}c_{1}:  \nonumber\\
& \ & \; \; \; + \delta_{1^\prime,2}\delta_{1,2^\prime}\left( n(1^\prime)+n(1)-2n(1^\prime)n(1) \right). \nonumber
\end{eqnarray}
In the  2-loop calculation we neglect terms with four or six fermionic operators on the rhs. in the following
since these terms would enter the calculation only in 3-loop order. We again need the commutator
\begin{eqnarray}
[:c_{1^{\prime}}^{\dagger}c_{1}:,:c_{2^{\prime}}^{\dagger}c_{2}c_{3^{\prime}}^{\dagger}c_{3}:] \; = \;  \;\:\;\;\:\;\;\:\;\;\:\;\;\:\; & \ & \\ 
 \delta_{1^{\prime},2}\delta_{1,2^{\prime}}(n(1^{\prime})-n(1)):c_{3^{\prime}}^{\dagger}c_{3}: & \ & \nonumber\\
 -\delta_{1^{\prime},2}\delta_{1,3^{\prime}}(n(1^{\prime})-n(1)):c_{2^{\prime}}^{\dagger}c_{3}: & \ & \nonumber\\
 -\delta_{1^{\prime},3}\delta_{1,2^{\prime}}(n(1^{\prime})-n(1)):c_{3^{\prime}}^{\dagger}c_{2}: & \ & \nonumber\\
 +\delta_{1^{\prime},3}\delta_{1,3^{\prime}}(n(1^{\prime})-n(1)):c_{2^{\prime}}^{\dagger}c_{2}: & \ &\nonumber
\end{eqnarray}
and the anticommutator
\begin{eqnarray}
\{:c_{1^{\prime}}^{\dagger}c_{1}:,:c_{2^{\prime}}^{\dagger}c_{2}c_{3^{\prime}}^{\dagger}c_{3}:\} \; = \;   \;\:\;\;\;\:\;\;\:\;\;\:\;\;\:\;\;\:\;\;\:\;\;\:\;\;\:\;\;\:\;\;\:\; \\ 
\delta_{1^{\prime},2}\delta_{1,2^{\prime}}(n(1^{\prime})+n(1)-2n(1^{\prime})n(1)):c_{3^{\prime}}^{\dagger}c_{3}: & \ & \nonumber\\
-\delta_{1^{\prime},2}\delta_{1,3^{\prime}}(n(1^{\prime})+n(1)-2n(1^{\prime})n(1)):c_{2^{\prime}}^{\dagger}c_{3}: & \ & \nonumber\\
-\delta_{1^{\prime},3}\delta_{1,2^{\prime}}(n(1^{\prime})+n(1)-2n(1^{\prime})n(1)):c_{3^{\prime}}^{\dagger}c_{2}: & \ & \nonumber\\
+\delta_{1^{\prime},3}\delta_{1,3^{\prime}}(n(1^{\prime})+n(1)-2n(1^{\prime})n(1)):c_{2^{\prime}}^{\dagger}c_{2}: & \ & \nonumber
\end{eqnarray}
for the further calculation.\\

Using the above relations the task of deriving the flow equations is reduced to simple but lengthy bookkeeping.
The resulting 2-loop equations are given in the following. In the diagonal part of the Hamiltonian only the splitting of the dot levels
due to the magnetic field is shifted
\begin{eqnarray}
\frac{dh}{dB} & = & \frac{1}{2} \sum\limits_{p,q} (n_f(p)+n_f(q)-2n_f(p)n_f(q)) \\
& \ & \;\; \times (\epsilon_p-\epsilon_q+h)(J^\perp_{pq})^2 +{\cal O}(J^4)\ .\nonumber
\end{eqnarray}
In the case of zero (initial) magnetic field the relation $J^\perp_{pq}=J^\perp_{qp}$ is fulfilled leading to $dh/dB=0$ and therefore
no additional magnetic field is generated.
In the interaction part we have to keep track of different scattering processes that lead to different flows of the running couplings though we
started with isotropic initial conditions.
\begin{widetext}
For the scattering of spin up electrons we find
\begin{eqnarray}
\frac{dJ^\uparrow_{pq}}{dB} & = & -(\epsilon_p-\epsilon_q)^2 J^\uparrow_{pq}
 + \frac{1}{2} \sum\limits_r (2(\epsilon_r-h)-(\epsilon_p+\epsilon_q)) J^\perp_{pr}J^\perp_{qr}(1-2n_f(r))\\
& \ & - \sum\limits_{r,s} ( n_f(r)+n_f(s)-2n_f(r)n_f(s) ) J^\perp_{rs} 
 \left( (\epsilon_p-\epsilon_q+2(\epsilon_r-\epsilon_s+h)) (K^\uparrow_{rs,pq} -K^\uparrow_{ps,rq}) \right.\nonumber\\
& \ & \left. \ \ \ -(\epsilon_p-\epsilon_q-2(\epsilon_r-\epsilon_s+h))
  (K^\uparrow_{rs,qp}-K^\uparrow_{qs,rp}) \right) +{\cal O}(J^4)\nonumber
\end{eqnarray}
and for spin down scattering
\begin{eqnarray}
\frac{dJ^\downarrow_{pq}}{dB} & = & -(\epsilon_p-\epsilon_q)^2 J^\downarrow_{pq}  + \frac{1}{2} \sum\limits_r (2(\epsilon_r+h)-(\epsilon_p+\epsilon_q)) J^\perp_{rp} J^\perp_{rq} (1-2n_f(r))\\
& \ & + \sum\limits_{r,s} ( n_f(r)+n_f(s)-2n_f(r)n_f(s) ) J^\perp_{rs} \left( (\epsilon_p-\epsilon_q+2(\epsilon_r-\epsilon_s+h)) (K^\downarrow_{rs,pq} 
-K^\downarrow_{rq,ps})\right. \nonumber\\
& \ & \left. \ \ \ - (\epsilon_p-\epsilon_q-2(\epsilon_r-\epsilon_s+h)) (K^\downarrow_{rs,qp} -K^\downarrow_{rp,qs}) \right)
 +{\cal O}(J^4)\ .\nonumber
\end{eqnarray}
The spin flip scattering is given by
\begin{eqnarray}
\frac{dJ^\perp_{pq}}{dB} & = & -(\epsilon_p-\epsilon_q+h)^2 J^\perp_{pq}  + \frac{1}{4} \sum\limits_r (1-2n_f(r)) 
\left( (2\epsilon_r-(\epsilon_p+\epsilon_q)+h) J^\perp_{rq} J^\uparrow_{pr} 
+ (2\epsilon_r-(\epsilon_p+\epsilon_q)-h) J^\perp_{pr}J^\downarrow_{qr} \right)\ \ \\
& \ & + \frac{1}{2} \sum\limits_{r,s} ( n_f(r)+n_f(s)-2n_f(r)n_f(s) ) 
( \epsilon_p-\epsilon_q+2(\epsilon_r-\epsilon_s) +h ) \nonumber\\
& \ & \ \ \ \times
\left( ( K^\uparrow_{pq,rs}-K^\uparrow_{rq,ps}) J^\uparrow_{sr} 
-(K^\downarrow_{pq,rs} -K^\downarrow_{ps,rq} ) J^\downarrow_{sr} \right) \nonumber\\
& \ & - \frac{1}{2}  \sum\limits_{r,s} ( n_f(r)+n_f(s)-2n_f(r)n_f(s) ) 
 ( \epsilon_p-\epsilon_q+2(\epsilon_r-\epsilon_s) -h ) K^\perp_{pq,rs} J^\perp_{sr} +{\cal O}(J^4)\ .\nonumber
\end{eqnarray}
\end{widetext}
The flow of the newly generated interactions is given by
\begin{eqnarray}
\frac{dK^\uparrow_{pq,rs}}{dB} & = & -( \epsilon_p-\epsilon_q+\epsilon_r-\epsilon_s+h )^2 K^\uparrow_{pq,rs} \\
& \ & +\frac{1}{4} (\epsilon_p-\epsilon_q-\epsilon_r+\epsilon_s+h) J^\perp_{pq}J^\uparrow_{rs} +{\cal O}(J^3)\nonumber
\end{eqnarray}
for the spin up plus spin flip scattering and 
\begin{eqnarray}
\frac{dK^\downarrow_{pq,rs}}{dB} & = & - (\epsilon_p-\epsilon_q+\epsilon_r-\epsilon_s+h)^2 K^\downarrow_{pq,rs} \\
& \ & -\frac{1}{4} (\epsilon_p-\epsilon_q-\epsilon_r+\epsilon_s+h) J^\perp_{pq}J^\downarrow_{rs} +{\cal O}(J^3)\nonumber
\end{eqnarray}
for spin down plus spin flip.
For double spin flip we find
\begin{eqnarray}
\frac{dK^\perp_{pq,rs}}{dB} & = & -( \epsilon_p-\epsilon_q+\epsilon_r-\epsilon_s )^2 K^\perp_{pq,rs} \\
& \ & -\frac{1}{2} (\epsilon_p-\epsilon_q-\epsilon_r+\epsilon_s+2h) J^\perp_{pq} J^\perp_{sr} +{\cal O}(J^3)\ .\nonumber
\end{eqnarray}

\section{Transformation of $S^{x/y}$}\label{XYZ}
For completeness we show the transformation of the spin operators perpendicular to the magnetic field and give the result for the corresponding
spin-spin correlation function and the response function.
The flow can be described using one set of running couplings for both $x$- and $y$-direction since the choice
of the basis in the $xy$-plane is arbitrary. Also the correlation and the response function are given by single functions for both directions.\\

We use the following ansatz for $x$-direction
\begin{eqnarray}
S^x(B) & = & h^{xy}(B) S^x + i\sum\limits_{p,q} \mu^\uparrow_{pq}(B) :f_{p\uparrow}^\dagger f_{q_\uparrow}: S^y \\
& \ & +i\sum\limits_{p,q} \mu^\downarrow_{pq}(B) :f_{p\downarrow}^\dagger f_{q_\downarrow}: S^y \nonumber\\
& \ & +\sum\limits_{p,q} \mu^z_{pq}(B) ( :f_{p\uparrow}^\dagger f_{q_\downarrow}: +  :f_{q\downarrow}^\dagger f_{p_\uparrow}:  ) S^z \nonumber
\end{eqnarray}
and $y$-direction
\begin{eqnarray}
S^y(B) & = & h^{xy}(B) S^y - i\sum\limits_{p,q} \mu^\uparrow_{pq}(B) :f_{p\uparrow}^\dagger f_{q_\uparrow}: S^x \\
& \ & -i\sum\limits_{p,q} \mu^\downarrow_{pq}(B) :f_{p\downarrow}^\dagger f_{q_\downarrow}: S^x \nonumber\\
& \ & -i\sum\limits_{p,q} \mu^z_{pq}(B) ( :f_{p\uparrow}^\dagger f_{q_\downarrow}: -  :f_{q\downarrow}^\dagger f_{p_\uparrow}:  ) S^z\ . \nonumber
\end{eqnarray}
The flow equation for the decay of the spin operator is given by
\begin{eqnarray}
\frac{dh^{xy}}{dB} & = & \frac{1}{4} \sum\limits_{p,q} (n_f(p)+n_f(q)-2n_f(p)n_f(q)) \\
& \ & \;\;\times (\epsilon_p-\epsilon_q)
(J^\uparrow_{pq}\mu^\uparrow_{qp}- J^\downarrow_{pq}\mu^\downarrow_{qp}) \nonumber \\
& \ & +\frac{1}{2} \sum\limits_{p,q} (n_f(p)+n_f(q)-2n_f(p)n_f(q)) \nonumber\\
& \ & \;\;\times (\epsilon_p-\epsilon_q+h)
J^\perp_{pq}\mu^z_{pq}\ . \nonumber
\end{eqnarray}
The flow of the newly generated operators is given by
\begin{eqnarray}
\frac{d\mu^\uparrow_{pq}}{dB} & = & \frac{h^{xy}}{2} (\epsilon_p-\epsilon_q) J^\uparrow_{pq} - \\
& \ & -\frac{1}{4} \sum\limits_r (1-2n_f(r)) (\epsilon_p-\epsilon_r+h) J^\perp_{pr} \mu^z_{qr} \nonumber\\
& \ & +\frac{1}{4} \sum\limits_r (1-2n_f(r)) (\epsilon_q-\epsilon_r+h) J^\perp_{qr} \mu^z_{pr}\nonumber
\end{eqnarray}
for the spin up component and
\begin{eqnarray}
\frac{d\mu^\downarrow_{pq}}{dB} & = & -\frac{h^{xy}}{2} (\epsilon_p-\epsilon_q) J^\downarrow_{pq} \\
& \ & -\frac{1}{4} \sum\limits_r (1-2n_f(r)) (\epsilon_r-\epsilon_q+h) J^\perp_{rq} \mu^z_{rp} \nonumber\\
& \ & +\frac{1}{4} \sum\limits_r (1-2n_f(r)) (\epsilon_r-\epsilon_p+h) J^\perp_{rp} \mu^z_{rq}\nonumber
\end{eqnarray}
for spin down. For the spin-flip component we find  
\begin{eqnarray}
\frac{d\mu^z_{pq}}{dB} & = & -\frac{h^{xy}}{2} (\epsilon_p-\epsilon_q+h) J^\perp_{pq} \\
& \ & -\frac{1}{4} \sum\limits_r (1-2n_f(r)) (\epsilon_r-\epsilon_q+h) J^\perp_{rq} \mu^\uparrow_{pr}\nonumber\\
& \ & -\frac{1}{4} \sum\limits_r (1-2n_f(r)) (\epsilon_p-\epsilon_r+h) J^\perp_{pr} \mu^\downarrow_{rq}\ .\nonumber
\end{eqnarray}
The correlation function is given by the lengthy formula
\begin{widetext}
\begin{eqnarray}
C^{xy}(\omega) & = & \frac{\pi(1+\text{sgn}(\tilde{h}))}{8} 
 \sum\limits_{p} 
( (\tilde{\mu}^\uparrow_{\epsilon_p,\epsilon_p+\omega-\tilde{h}})^2 +
(\tilde{\mu}^\downarrow_{\epsilon_p,\epsilon_p+\omega-\tilde{h}})^2 ) 
 n_f(\epsilon_p)(1-n_f(\epsilon_p+\omega-\tilde{h}))\nonumber\\
& \ & +\frac{\pi(1+\text{sgn}(\tilde{h}))}{8} 
 \sum\limits_{p} ( (\tilde{\mu}^\uparrow_{\epsilon_p,\epsilon_p-\omega-\tilde{h}})^2
+(\tilde{\mu}^\uparrow_{\epsilon_p,\epsilon_p-\omega-\tilde{h}})^2 ) 
n_f(\epsilon_p)(1-n_f(\epsilon_p-\omega-\tilde{h}))\nonumber\\
& \ & +\frac{\pi(1-\text{sgn}(\tilde{h}))}{8} 
 \sum\limits_{p} ( (\tilde{\mu}^\uparrow_{\epsilon_p,\epsilon_p+\omega+\tilde{h}})^2 
+(\tilde{\mu}^\downarrow_{\epsilon_p,\epsilon_p+\omega+\tilde{h}})^2 ) 
n_f(\epsilon_p)(1-n_f(\epsilon_p+\omega+\tilde{h}))\nonumber\\
& \ & +\frac{\pi(1-\text{sgn}(\tilde{h}))}{8} 
 \sum\limits_{p} ( (\tilde{\mu}^\uparrow_{\epsilon_p,\epsilon_p-\omega+\tilde{h}})^2
+(\tilde{\mu}^\downarrow_{\epsilon_p,\epsilon_p-\omega+\tilde{h}})^2 ) 
n_f(\epsilon_p)(1-n_f(\epsilon_p-\omega+\tilde{h}))\nonumber\\
& \ & +\frac{\pi}{4} \sum\limits_p ( \tilde{\mu}^z_{\epsilon_p,\epsilon_p+\omega} )^2 (n_f(\epsilon_p) (1-n_f(\epsilon_p+\omega))
+ n_f(\epsilon_p+\omega) (1-n_f(\epsilon_p)) )
\nonumber\\
& \ & +\frac{\pi}{4} \sum\limits_p ( \tilde{\mu}^z_{\epsilon_p,\epsilon_p-\omega} )^2  ( n_f(\epsilon_p) (1-n_f(\epsilon_p-\omega))
+n_f(\epsilon_p-\omega) (1-n_f(\epsilon_p)))\ .
\end{eqnarray}
For the imaginary part of the response function we find
\begin{eqnarray}
\chi_{xy}^{\prime\prime}(\omega) & = & \frac{\pi(1+\text{sgn}(\tilde{h}))}{8}
\sum\limits_{p} 
( (\tilde{\mu}^\uparrow_{\epsilon_p,\epsilon_p+\omega-\tilde{h}})^2 +
(\tilde{\mu}^\downarrow_{\epsilon_p,\epsilon_p+\omega-\tilde{h}})^2 ) 
n_f(\epsilon_p)(1-n_f(\epsilon_p+\omega-\tilde{h}))\nonumber\\
& \ & -\frac{\pi(1+\text{sgn}(\tilde{h}))}{8} 
\sum\limits_{p} ( (\tilde{\mu}^\uparrow_{\epsilon_p,\epsilon_p-\omega-\tilde{h}})^2
+(\tilde{\mu}^\uparrow_{\epsilon_p,\epsilon_p-\omega-\tilde{h}})^2 ) 
n_f(\epsilon_p)(1-n_f(\epsilon_p-\omega-\tilde{h}))\nonumber\\
& \ & +\frac{\pi(1-\text{sgn}(\tilde{h}))}{8} 
 \sum\limits_{p} ( (\tilde{\mu}^\uparrow_{\epsilon_p,\epsilon_p+\omega+\tilde{h}})^2 
+(\tilde{\mu}^\downarrow_{\epsilon_p,\epsilon_p+\omega+\tilde{h}})^2 ) 
n_f(\epsilon_p)(1-n_f(\epsilon_p+\omega+\tilde{h}))\nonumber\\
& \ & -\frac{\pi(1-\text{sgn}(\tilde{h}))}{8} 
 \sum\limits_{p} ( (\tilde{\mu}^\uparrow_{\epsilon_p,\epsilon_p-\omega+\tilde{h}})^2
+(\tilde{\mu}^\downarrow_{\epsilon_p,\epsilon_p-\omega+\tilde{h}})^2 ) 
n_f(\epsilon_p)(1-n_f(\epsilon_p-\omega+\tilde{h}))\nonumber\\
& \ & +\frac{\pi}{4} \sum\limits_p ( \tilde{\mu}^z_{\epsilon_p,\epsilon_p+\omega} )^2 (n_f(\epsilon_p) (1-n_f(\epsilon_p+\omega))
- n_f(\epsilon_p+\omega) (1-n_f(\epsilon_p)) )
\nonumber\\
& \ & -\frac{\pi}{4} \sum\limits_p ( \tilde{\mu}^z_{\epsilon_p,\epsilon_p-\omega} )^2  ( n_f(\epsilon_p) (1-n_f(\epsilon_p-\omega))
-n_f(\epsilon_p-\omega) (1-n_f(\epsilon_p)))\ .
\end{eqnarray}
Note that the equations above are identical to the transformation of the $S^z$ operator at zero magnetic field.
\end{widetext}


\begin{thebibliography}{64}
\bibitem[{not()}]{note_normal_order}
\bibinfo{note}{It is sufficient to normal order with respect to the
  non-interacting ground state/thermal state since the corrections enter in ${\cal O}(J^4)$
  and we will derive the flow equations only in ${\cal O}(J^3)$.}
\end{thebibliography}
\end{document}